\documentclass[submission,copyright,creativecommons]{eptcs}

\providecommand\publicationstatus{}

\usepackage{graphicx}
\usepackage[utf8]{inputenc}
\usepackage[T1]{fontenc}

\usepackage{cite}

\usepackage{booktabs}

\providecommand{\event}{ThEdu'21 Proceedings} 
\usepackage{underscore}           

\usepackage{isabelle, isabellesym}
\isabellestyle{it} 
\newcommand{\DefineSnippet}[2]{\expandafter\newcommand\csname snippet--#1\endcsname{#2}}
\DefineSnippet{theory:Pure-I-0}{%
\isadelimtheory
\endisadelimtheory
\isatagtheory
\isacommand{theory}\isamarkupfalse%
\ Pure{\isacharunderscore}{\kern0pt}I\ \isakeyword{imports}\ Pure\ \isakeyword{begin}%
\endisatagtheory
{\isafoldtheory}%
\isadelimtheory
\endisadelimtheory
}
\DefineSnippet{typedecl:f7b94bb19014cba6-0}{%
\isacommand{typedecl}\isamarkupfalse%
\ bool%
}
\DefineSnippet{judgment:Trueprop-0}{%
\isacommand{judgment}\isamarkupfalse%
\ Trueprop\ {\isacharcolon}{\kern0pt}{\isacharcolon}{\kern0pt}\ {\isacartoucheopen}bool\ {\isasymRightarrow}\ prop{\isacartoucheclose}\ {\isacharparenleft}{\kern0pt}{\isacartoucheopen}{\isacharunderscore}{\kern0pt}{\isacartoucheclose}{\isacharparenright}{\kern0pt}%
}
\DefineSnippet{axiomatization:Imp-0}{%
\isacommand{axiomatization}\isamarkupfalse%
\ Imp\ {\isacharparenleft}{\kern0pt}\isakeyword{infixr}\ {\isacartoucheopen}{\isasymlongrightarrow}{\isacartoucheclose}\ {\isadigit{3}}{\isacharparenright}{\kern0pt}\isanewline
}
\DefineSnippet{axiomatization:Imp-1}{%
\ \ \isakeyword{where}\ Imp{\isacharunderscore}{\kern0pt}I\ {\isacharbrackleft}{\kern0pt}intro{\isacharbrackright}{\kern0pt}{\isacharcolon}{\kern0pt}\ {\isacartoucheopen}{\isacharparenleft}{\kern0pt}p\ {\isasymLongrightarrow}\ q{\isacharparenright}{\kern0pt}\ {\isasymLongrightarrow}\ p\ {\isasymlongrightarrow}\ q{\isacartoucheclose}\isanewline
}
\DefineSnippet{axiomatization:Imp-2}{%
\ \ \ \ \isakeyword{and}\ Imp{\isacharunderscore}{\kern0pt}E\ {\isacharbrackleft}{\kern0pt}elim{\isacharbrackright}{\kern0pt}{\isacharcolon}{\kern0pt}\ {\isacartoucheopen}p\ {\isasymlongrightarrow}\ q\ {\isasymLongrightarrow}\ p\ {\isasymLongrightarrow}\ q{\isacartoucheclose}%
}
\DefineSnippet{axiomatization:Dis-0}{%
\isacommand{axiomatization}\isamarkupfalse%
\ Dis\ {\isacharparenleft}{\kern0pt}\isakeyword{infixr}\ {\isacartoucheopen}{\isasymor}{\isacartoucheclose}\ {\isadigit{4}}{\isacharparenright}{\kern0pt}\isanewline
}
\DefineSnippet{axiomatization:Dis-1}{%
\ \ \isakeyword{where}\ Dis{\isacharunderscore}{\kern0pt}E\ {\isacharbrackleft}{\kern0pt}elim{\isacharbrackright}{\kern0pt}{\isacharcolon}{\kern0pt}\ {\isacartoucheopen}p\ {\isasymor}\ q\ {\isasymLongrightarrow}\ {\isacharparenleft}{\kern0pt}p\ {\isasymLongrightarrow}\ r{\isacharparenright}{\kern0pt}\ {\isasymLongrightarrow}\ {\isacharparenleft}{\kern0pt}q\ {\isasymLongrightarrow}\ r{\isacharparenright}{\kern0pt}\ {\isasymLongrightarrow}\ r{\isacartoucheclose}\isanewline
}
\DefineSnippet{axiomatization:Dis-2}{%
\ \ \ \ \isakeyword{and}\ Dis{\isacharunderscore}{\kern0pt}I{\isadigit{1}}\ {\isacharbrackleft}{\kern0pt}intro{\isacharbrackright}{\kern0pt}{\isacharcolon}{\kern0pt}\ {\isacartoucheopen}p\ {\isasymLongrightarrow}\ p\ {\isasymor}\ q{\isacartoucheclose}\isanewline
}
\DefineSnippet{axiomatization:Dis-3}{%
\ \ \ \ \isakeyword{and}\ Dis{\isacharunderscore}{\kern0pt}I{\isadigit{2}}\ {\isacharbrackleft}{\kern0pt}intro{\isacharbrackright}{\kern0pt}{\isacharcolon}{\kern0pt}\ {\isacartoucheopen}q\ {\isasymLongrightarrow}\ p\ {\isasymor}\ q{\isacartoucheclose}%
}
\DefineSnippet{axiomatization:Con-0}{%
\isacommand{axiomatization}\isamarkupfalse%
\ Con\ {\isacharparenleft}{\kern0pt}\isakeyword{infixr}\ {\isacartoucheopen}{\isasymand}{\isacartoucheclose}\ {\isadigit{5}}{\isacharparenright}{\kern0pt}\isanewline
}
\DefineSnippet{axiomatization:Con-1}{%
\ \ \isakeyword{where}\ Con{\isacharunderscore}{\kern0pt}I\ {\isacharbrackleft}{\kern0pt}intro{\isacharbrackright}{\kern0pt}{\isacharcolon}{\kern0pt}\ {\isacartoucheopen}p\ {\isasymLongrightarrow}\ q\ {\isasymLongrightarrow}\ p\ {\isasymand}\ q{\isacartoucheclose}\isanewline
}
\DefineSnippet{axiomatization:Con-2}{%
\ \ \ \ \isakeyword{and}\ Con{\isacharunderscore}{\kern0pt}E{\isadigit{1}}\ {\isacharbrackleft}{\kern0pt}elim{\isacharbrackright}{\kern0pt}{\isacharcolon}{\kern0pt}\ {\isacartoucheopen}p\ {\isasymand}\ q\ {\isasymLongrightarrow}\ p{\isacartoucheclose}\isanewline
}
\DefineSnippet{axiomatization:Con-3}{%
\ \ \ \ \isakeyword{and}\ Con{\isacharunderscore}{\kern0pt}E{\isadigit{2}}\ {\isacharbrackleft}{\kern0pt}elim{\isacharbrackright}{\kern0pt}{\isacharcolon}{\kern0pt}\ {\isacartoucheopen}p\ {\isasymand}\ q\ {\isasymLongrightarrow}\ q{\isacartoucheclose}%
}
\DefineSnippet{axiomatization:Falsity-0}{%
\isacommand{axiomatization}\isamarkupfalse%
\ Falsity\ {\isacharparenleft}{\kern0pt}{\isacartoucheopen}{\isasymbottom}{\isacartoucheclose}{\isacharparenright}{\kern0pt}\isanewline
}
\DefineSnippet{axiomatization:Falsity-1}{%
\ \ \isakeyword{where}\ Falsity{\isacharunderscore}{\kern0pt}E\ {\isacharbrackleft}{\kern0pt}elim{\isacharbrackright}{\kern0pt}{\isacharcolon}{\kern0pt}\ {\isacartoucheopen}{\isasymbottom}\ {\isasymLongrightarrow}\ p{\isacartoucheclose}%
}
\DefineSnippet{definition:Truth-0}{%
\isacommand{definition}\isamarkupfalse%
\ Truth\ {\isacharparenleft}{\kern0pt}{\isacartoucheopen}{\isasymtop}{\isacartoucheclose}{\isacharparenright}{\kern0pt}\ \isakeyword{where}\ {\isacartoucheopen}{\isasymtop}\ {\isasymequiv}\ {\isasymbottom}\ {\isasymlongrightarrow}\ {\isasymbottom}{\isacartoucheclose}%
}
\DefineSnippet{theorem:Truth-I-0}{%
\isacommand{theorem}\isamarkupfalse%
\ Truth{\isacharunderscore}{\kern0pt}I\ {\isacharbrackleft}{\kern0pt}intro{\isacharbrackright}{\kern0pt}{\isacharcolon}{\kern0pt}\ {\isacartoucheopen}{\isasymtop}{\isacartoucheclose}\isanewline
}
\DefineSnippet{theorem:Truth-I-1}{%
\isadelimproof
\ \ %
\endisadelimproof
\isatagproof
\isacommand{unfolding}\isamarkupfalse%
\ Truth{\isacharunderscore}{\kern0pt}def\ \isacommand{{\isachardot}{\kern0pt}{\isachardot}{\kern0pt}}\isamarkupfalse%
\endisatagproof
{\isafoldproof}%
\isadelimproof
\endisadelimproof
}
\DefineSnippet{definition:Neg-0}{%
\isacommand{definition}\isamarkupfalse%
\ Neg\ {\isacharparenleft}{\kern0pt}{\isacartoucheopen}{\isasymnot}\ {\isacharunderscore}{\kern0pt}{\isacartoucheclose}\ {\isacharbrackleft}{\kern0pt}{\isadigit{6}}{\isacharbrackright}{\kern0pt}\ {\isadigit{6}}{\isacharparenright}{\kern0pt}\ \isakeyword{where}\ {\isacartoucheopen}{\isasymnot}\ p\ {\isasymequiv}\ p\ {\isasymlongrightarrow}\ {\isasymbottom}{\isacartoucheclose}%
}
\DefineSnippet{theorem:Neg-I-0}{%
\isacommand{theorem}\isamarkupfalse%
\ Neg{\isacharunderscore}{\kern0pt}I\ {\isacharbrackleft}{\kern0pt}intro{\isacharbrackright}{\kern0pt}{\isacharcolon}{\kern0pt}\ {\isacartoucheopen}{\isacharparenleft}{\kern0pt}p\ {\isasymLongrightarrow}\ {\isasymbottom}{\isacharparenright}{\kern0pt}\ {\isasymLongrightarrow}\ {\isasymnot}\ p{\isacartoucheclose}\isanewline
}
\DefineSnippet{theorem:Neg-I-1}{%
\isadelimproof
\ \ %
\endisadelimproof
\isatagproof
\isacommand{unfolding}\isamarkupfalse%
\ Neg{\isacharunderscore}{\kern0pt}def\ \isacommand{{\isachardot}{\kern0pt}{\isachardot}{\kern0pt}}\isamarkupfalse%
\endisatagproof
{\isafoldproof}%
\isadelimproof
\endisadelimproof
}
\DefineSnippet{theorem:Neg-E-0}{%
\isacommand{theorem}\isamarkupfalse%
\ Neg{\isacharunderscore}{\kern0pt}E\ {\isacharbrackleft}{\kern0pt}elim{\isacharbrackright}{\kern0pt}{\isacharcolon}{\kern0pt}\ {\isacartoucheopen}{\isasymnot}\ p\ {\isasymLongrightarrow}\ p\ {\isasymLongrightarrow}\ q{\isacartoucheclose}\isanewline
}
\DefineSnippet{theorem:Neg-E-1}{%
\isadelimproof
\ \ %
\endisadelimproof
\isatagproof
\isacommand{unfolding}\isamarkupfalse%
\ Neg{\isacharunderscore}{\kern0pt}def\isanewline
}
\DefineSnippet{theorem:Neg-E-2}{%
\isacommand{proof}\isamarkupfalse%
\ {\isacharminus}{\kern0pt}\isanewline
}
\DefineSnippet{theorem:Neg-E-3}{%
\ \ \isacommand{assume}\isamarkupfalse%
\ {\isacartoucheopen}p\ {\isasymlongrightarrow}\ {\isasymbottom}{\isacartoucheclose}\ \isakeyword{and}\ {\isacartoucheopen}p{\isacartoucheclose}\isanewline
}
\DefineSnippet{theorem:Neg-E-4}{%
\ \ \isacommand{then}\isamarkupfalse%
\ \isacommand{have}\isamarkupfalse%
\ {\isacartoucheopen}{\isasymbottom}{\isacartoucheclose}\ \isacommand{{\isachardot}{\kern0pt}{\isachardot}{\kern0pt}}\isamarkupfalse%
\isanewline
}
\DefineSnippet{theorem:Neg-E-5}{%
\ \ \isacommand{then}\isamarkupfalse%
\ \isacommand{show}\isamarkupfalse%
\ {\isacartoucheopen}q{\isacartoucheclose}\ \isacommand{{\isachardot}{\kern0pt}{\isachardot}{\kern0pt}}\isamarkupfalse%
\isanewline
}
\DefineSnippet{theorem:Neg-E-6}{%
\isacommand{qed}\isamarkupfalse%
\endisatagproof
{\isafoldproof}%
\isadelimproof
\endisadelimproof
}
\DefineSnippet{definition:Iff-0}{%
\isacommand{definition}\isamarkupfalse%
\ Iff\ {\isacharparenleft}{\kern0pt}\isakeyword{infixr}\ {\isacartoucheopen}{\isasymlongleftrightarrow}{\isacartoucheclose}\ {\isadigit{3}}{\isacharparenright}{\kern0pt}\ \isakeyword{where}\ {\isacartoucheopen}p\ {\isasymlongleftrightarrow}\ q\ {\isasymequiv}\ {\isacharparenleft}{\kern0pt}p\ {\isasymlongrightarrow}\ q{\isacharparenright}{\kern0pt}\ {\isasymand}\ {\isacharparenleft}{\kern0pt}q\ {\isasymlongrightarrow}\ p{\isacharparenright}{\kern0pt}{\isacartoucheclose}%
}
\DefineSnippet{theorem:Iff-I-0}{%
\isacommand{theorem}\isamarkupfalse%
\ Iff{\isacharunderscore}{\kern0pt}I\ {\isacharbrackleft}{\kern0pt}intro{\isacharbrackright}{\kern0pt}{\isacharcolon}{\kern0pt}\ {\isacartoucheopen}{\isacharparenleft}{\kern0pt}p\ {\isasymLongrightarrow}\ q{\isacharparenright}{\kern0pt}\ {\isasymLongrightarrow}\ {\isacharparenleft}{\kern0pt}q\ {\isasymLongrightarrow}\ p{\isacharparenright}{\kern0pt}\ {\isasymLongrightarrow}\ p\ {\isasymlongleftrightarrow}\ q{\isacartoucheclose}\isanewline
}
\DefineSnippet{theorem:Iff-I-1}{%
\isadelimproof
\ \ %
\endisadelimproof
\isatagproof
\isacommand{unfolding}\isamarkupfalse%
\ Iff{\isacharunderscore}{\kern0pt}def\isanewline
}
\DefineSnippet{theorem:Iff-I-2}{%
\isacommand{proof}\isamarkupfalse%
\ {\isacharminus}{\kern0pt}\isanewline
}
\DefineSnippet{theorem:Iff-I-3}{%
\ \ \isacommand{assume}\isamarkupfalse%
\ {\isacartoucheopen}p\ {\isasymLongrightarrow}\ q{\isacartoucheclose}\ \isakeyword{and}\ {\isacartoucheopen}q\ {\isasymLongrightarrow}\ p{\isacartoucheclose}\isanewline
}
\DefineSnippet{theorem:Iff-I-4}{%
\ \ \isacommand{from}\isamarkupfalse%
\ {\isacartoucheopen}p\ {\isasymLongrightarrow}\ q{\isacartoucheclose}\ \isacommand{have}\isamarkupfalse%
\ {\isacartoucheopen}p\ {\isasymlongrightarrow}\ q{\isacartoucheclose}\ \isacommand{{\isachardot}{\kern0pt}{\isachardot}{\kern0pt}}\isamarkupfalse%
\isanewline
}
\DefineSnippet{theorem:Iff-I-5}{%
\ \ \isacommand{from}\isamarkupfalse%
\ {\isacartoucheopen}q\ {\isasymLongrightarrow}\ p{\isacartoucheclose}\ \isacommand{have}\isamarkupfalse%
\ {\isacartoucheopen}q\ {\isasymlongrightarrow}\ p{\isacartoucheclose}\ \isacommand{{\isachardot}{\kern0pt}{\isachardot}{\kern0pt}}\isamarkupfalse%
\isanewline
}
\DefineSnippet{theorem:Iff-I-6}{%
\ \ \isacommand{from}\isamarkupfalse%
\ {\isacartoucheopen}p\ {\isasymlongrightarrow}\ q{\isacartoucheclose}\ \isakeyword{and}\ {\isacartoucheopen}q\ {\isasymlongrightarrow}\ p{\isacartoucheclose}\ \isacommand{show}\isamarkupfalse%
\ {\isacartoucheopen}{\isacharparenleft}{\kern0pt}p\ {\isasymlongrightarrow}\ q{\isacharparenright}{\kern0pt}\ {\isasymand}\ {\isacharparenleft}{\kern0pt}q\ {\isasymlongrightarrow}\ p{\isacharparenright}{\kern0pt}{\isacartoucheclose}\ \isacommand{{\isachardot}{\kern0pt}{\isachardot}{\kern0pt}}\isamarkupfalse%
\isanewline
}
\DefineSnippet{theorem:Iff-I-7}{%
\isacommand{qed}\isamarkupfalse%
\endisatagproof
{\isafoldproof}%
\isadelimproof
\endisadelimproof
}
\DefineSnippet{theorem:Iff-E1-0}{%
\isacommand{theorem}\isamarkupfalse%
\ Iff{\isacharunderscore}{\kern0pt}E{\isadigit{1}}\ {\isacharbrackleft}{\kern0pt}elim{\isacharbrackright}{\kern0pt}{\isacharcolon}{\kern0pt}\ {\isacartoucheopen}p\ {\isasymlongleftrightarrow}\ q\ {\isasymLongrightarrow}\ p\ {\isasymLongrightarrow}\ q{\isacartoucheclose}\isanewline
}
\DefineSnippet{theorem:Iff-E1-1}{%
\isadelimproof
\ \ %
\endisadelimproof
\isatagproof
\isacommand{unfolding}\isamarkupfalse%
\ Iff{\isacharunderscore}{\kern0pt}def\isanewline
}
\DefineSnippet{theorem:Iff-E1-2}{%
\isacommand{proof}\isamarkupfalse%
\ {\isacharminus}{\kern0pt}\isanewline
}
\DefineSnippet{theorem:Iff-E1-3}{%
\ \ \isacommand{assume}\isamarkupfalse%
\ {\isacartoucheopen}{\isacharparenleft}{\kern0pt}p\ {\isasymlongrightarrow}\ q{\isacharparenright}{\kern0pt}\ {\isasymand}\ {\isacharparenleft}{\kern0pt}q\ {\isasymlongrightarrow}\ p{\isacharparenright}{\kern0pt}{\isacartoucheclose}\isanewline
}
\DefineSnippet{theorem:Iff-E1-4}{%
\ \ \isacommand{then}\isamarkupfalse%
\ \isacommand{have}\isamarkupfalse%
\ {\isacartoucheopen}p\ {\isasymlongrightarrow}\ q{\isacartoucheclose}\ \isacommand{{\isachardot}{\kern0pt}{\isachardot}{\kern0pt}}\isamarkupfalse%
\isanewline
}
\DefineSnippet{theorem:Iff-E1-5}{%
\ \ \isacommand{then}\isamarkupfalse%
\ \isacommand{show}\isamarkupfalse%
\ {\isacartoucheopen}p\ {\isasymLongrightarrow}\ q{\isacartoucheclose}\ \isacommand{{\isachardot}{\kern0pt}{\isachardot}{\kern0pt}}\isamarkupfalse%
\isanewline
}
\DefineSnippet{theorem:Iff-E1-6}{%
\isacommand{qed}\isamarkupfalse%
\endisatagproof
{\isafoldproof}%
\isadelimproof
\endisadelimproof
}
\DefineSnippet{theorem:Iff-E2-0}{%
\isacommand{theorem}\isamarkupfalse%
\ Iff{\isacharunderscore}{\kern0pt}E{\isadigit{2}}\ {\isacharbrackleft}{\kern0pt}elim{\isacharbrackright}{\kern0pt}{\isacharcolon}{\kern0pt}\ {\isacartoucheopen}p\ {\isasymlongleftrightarrow}\ q\ {\isasymLongrightarrow}\ q\ {\isasymLongrightarrow}\ p{\isacartoucheclose}\isanewline
}
\DefineSnippet{theorem:Iff-E2-1}{%
\isadelimproof
\ \ %
\endisadelimproof
\isatagproof
\isacommand{unfolding}\isamarkupfalse%
\ Iff{\isacharunderscore}{\kern0pt}def\isanewline
}
\DefineSnippet{theorem:Iff-E2-2}{%
\isacommand{proof}\isamarkupfalse%
\ {\isacharminus}{\kern0pt}\isanewline
}
\DefineSnippet{theorem:Iff-E2-3}{%
\ \ \isacommand{assume}\isamarkupfalse%
\ {\isacartoucheopen}{\isacharparenleft}{\kern0pt}p\ {\isasymlongrightarrow}\ q{\isacharparenright}{\kern0pt}\ {\isasymand}\ {\isacharparenleft}{\kern0pt}q\ {\isasymlongrightarrow}\ p{\isacharparenright}{\kern0pt}{\isacartoucheclose}\isanewline
}
\DefineSnippet{theorem:Iff-E2-4}{%
\ \ \isacommand{then}\isamarkupfalse%
\ \isacommand{have}\isamarkupfalse%
\ {\isacartoucheopen}q\ {\isasymlongrightarrow}\ p{\isacartoucheclose}\ \isacommand{{\isachardot}{\kern0pt}{\isachardot}{\kern0pt}}\isamarkupfalse%
\isanewline
}
\DefineSnippet{theorem:Iff-E2-5}{%
\ \ \isacommand{then}\isamarkupfalse%
\ \isacommand{show}\isamarkupfalse%
\ {\isacartoucheopen}q\ {\isasymLongrightarrow}\ p{\isacartoucheclose}\ \isacommand{{\isachardot}{\kern0pt}{\isachardot}{\kern0pt}}\isamarkupfalse%
\isanewline
}
\DefineSnippet{theorem:Iff-E2-6}{%
\isacommand{qed}\isamarkupfalse%
\endisatagproof
{\isafoldproof}%
\isadelimproof
\endisadelimproof
}
\DefineSnippet{proposition:c034bbb10ac00aeb-0}{%
\isacommand{proposition}\isamarkupfalse%
\ {\isacartoucheopen}p\ {\isasymlongrightarrow}\ {\isasymnot}\ {\isasymnot}\ p{\isacartoucheclose}\isanewline
}
\DefineSnippet{proposition:c034bbb10ac00aeb-1}{%
\isadelimproof
\endisadelimproof
\isatagproof
\isacommand{proof}\isamarkupfalse%
\isanewline
}
\DefineSnippet{proposition:c034bbb10ac00aeb-2}{%
\ \ \isacommand{assume}\isamarkupfalse%
\ {\isacartoucheopen}p{\isacartoucheclose}\isanewline
}
\DefineSnippet{proposition:c034bbb10ac00aeb-3}{%
\ \ \isacommand{show}\isamarkupfalse%
\ {\isacartoucheopen}{\isasymnot}\ {\isasymnot}\ p{\isacartoucheclose}\isanewline
}
\DefineSnippet{proposition:c034bbb10ac00aeb-4}{%
\ \ \isacommand{proof}\isamarkupfalse%
\isanewline
}
\DefineSnippet{proposition:c034bbb10ac00aeb-5}{%
\ \ \ \ \isacommand{assume}\isamarkupfalse%
\ {\isacartoucheopen}{\isasymnot}\ p{\isacartoucheclose}\isanewline
}
\DefineSnippet{proposition:c034bbb10ac00aeb-6}{%
\ \ \ \ \isacommand{from}\isamarkupfalse%
\ {\isacartoucheopen}{\isasymnot}\ p{\isacartoucheclose}\ \isakeyword{and}\ {\isacartoucheopen}p{\isacartoucheclose}\ \isacommand{show}\isamarkupfalse%
\ {\isacartoucheopen}{\isasymbottom}{\isacartoucheclose}\ \isacommand{{\isachardot}{\kern0pt}{\isachardot}{\kern0pt}}\isamarkupfalse%
\isanewline
}
\DefineSnippet{proposition:c034bbb10ac00aeb-7}{%
\ \ \isacommand{qed}\isamarkupfalse%
\isanewline
}
\DefineSnippet{proposition:c034bbb10ac00aeb-8}{%
\isacommand{qed}\isamarkupfalse%
\endisatagproof
{\isafoldproof}%
\isadelimproof
\endisadelimproof
}
\DefineSnippet{proposition:88d0827f917d9c91-0}{%
\isacommand{proposition}\isamarkupfalse%
\ {\isacartoucheopen}{\isacharparenleft}{\kern0pt}p\ {\isasymlongrightarrow}\ q{\isacharparenright}{\kern0pt}\ {\isasymand}\ {\isasymnot}\ q\ {\isasymlongrightarrow}\ {\isasymnot}\ p{\isacartoucheclose}\isanewline
}
\DefineSnippet{proposition:88d0827f917d9c91-1}{%
\isadelimproof
\endisadelimproof
\isatagproof
\isacommand{proof}\isamarkupfalse%
\isanewline
}
\DefineSnippet{proposition:88d0827f917d9c91-2}{%
\ \ \isacommand{assume}\isamarkupfalse%
\ {\isacartoucheopen}{\isacharparenleft}{\kern0pt}p\ {\isasymlongrightarrow}\ q{\isacharparenright}{\kern0pt}\ {\isasymand}\ {\isasymnot}\ q{\isacartoucheclose}\isanewline
}
\DefineSnippet{proposition:88d0827f917d9c91-3}{%
\ \ \isacommand{show}\isamarkupfalse%
\ {\isacartoucheopen}{\isasymnot}\ p{\isacartoucheclose}\isanewline
}
\DefineSnippet{proposition:88d0827f917d9c91-4}{%
\ \ \isacommand{proof}\isamarkupfalse%
\isanewline
}
\DefineSnippet{proposition:88d0827f917d9c91-5}{%
\ \ \ \ \isacommand{assume}\isamarkupfalse%
\ {\isacartoucheopen}p{\isacartoucheclose}\isanewline
}
\DefineSnippet{proposition:88d0827f917d9c91-6}{%
\ \ \ \ \isacommand{from}\isamarkupfalse%
\ {\isacartoucheopen}{\isacharparenleft}{\kern0pt}p\ {\isasymlongrightarrow}\ q{\isacharparenright}{\kern0pt}\ {\isasymand}\ {\isasymnot}\ q{\isacartoucheclose}\ \isacommand{have}\isamarkupfalse%
\ {\isacartoucheopen}p\ {\isasymlongrightarrow}\ q{\isacartoucheclose}\ \isacommand{{\isachardot}{\kern0pt}{\isachardot}{\kern0pt}}\isamarkupfalse%
\isanewline
}
\DefineSnippet{proposition:88d0827f917d9c91-7}{%
\ \ \ \ \isacommand{from}\isamarkupfalse%
\ {\isacartoucheopen}p\ {\isasymlongrightarrow}\ q{\isacartoucheclose}\ \isakeyword{and}\ {\isacartoucheopen}p{\isacartoucheclose}\ \isacommand{have}\isamarkupfalse%
\ {\isacartoucheopen}q{\isacartoucheclose}\ \isacommand{{\isachardot}{\kern0pt}{\isachardot}{\kern0pt}}\isamarkupfalse%
\isanewline
}
\DefineSnippet{proposition:88d0827f917d9c91-8}{%
\ \ \ \ \isacommand{from}\isamarkupfalse%
\ {\isacartoucheopen}{\isacharparenleft}{\kern0pt}p\ {\isasymlongrightarrow}\ q{\isacharparenright}{\kern0pt}\ {\isasymand}\ {\isasymnot}\ q{\isacartoucheclose}\ \isacommand{have}\isamarkupfalse%
\ {\isacartoucheopen}{\isasymnot}\ q{\isacartoucheclose}\ \isacommand{{\isachardot}{\kern0pt}{\isachardot}{\kern0pt}}\isamarkupfalse%
\isanewline
}
\DefineSnippet{proposition:88d0827f917d9c91-9}{%
\ \ \ \ \isacommand{from}\isamarkupfalse%
\ {\isacartoucheopen}{\isasymnot}\ q{\isacartoucheclose}\ \isakeyword{and}\ {\isacartoucheopen}q{\isacartoucheclose}\ \isacommand{show}\isamarkupfalse%
\ {\isacartoucheopen}{\isasymbottom}{\isacartoucheclose}\ \isacommand{{\isachardot}{\kern0pt}{\isachardot}{\kern0pt}}\isamarkupfalse%
\isanewline
}
\DefineSnippet{proposition:88d0827f917d9c91-10}{%
\ \ \isacommand{qed}\isamarkupfalse%
\isanewline
}
\DefineSnippet{proposition:88d0827f917d9c91-11}{%
\isacommand{qed}\isamarkupfalse%
\endisatagproof
{\isafoldproof}%
\isadelimproof
\endisadelimproof
}
\DefineSnippet{proposition:ebcf409125ae6be5-0}{%
\isacommand{proposition}\isamarkupfalse%
\ {\isacartoucheopen}{\isacharparenleft}{\kern0pt}p\ {\isasymlongleftrightarrow}\ q{\isacharparenright}{\kern0pt}\ {\isasymlongleftrightarrow}\ q\ {\isasymlongleftrightarrow}\ p{\isacartoucheclose}\isanewline
}
\DefineSnippet{proposition:ebcf409125ae6be5-1}{%
\isadelimproof
\endisadelimproof
\isatagproof
\isacommand{proof}\isamarkupfalse%
\isanewline
}
\DefineSnippet{proposition:ebcf409125ae6be5-2}{%
\ \ \isacommand{assume}\isamarkupfalse%
\ {\isacartoucheopen}p\ {\isasymlongleftrightarrow}\ q{\isacartoucheclose}\isanewline
}
\DefineSnippet{proposition:ebcf409125ae6be5-3}{%
\ \ \isacommand{show}\isamarkupfalse%
\ {\isacartoucheopen}q\ {\isasymlongleftrightarrow}\ p{\isacartoucheclose}\isanewline
}
\DefineSnippet{proposition:ebcf409125ae6be5-4}{%
\ \ \isacommand{proof}\isamarkupfalse%
\isanewline
}
\DefineSnippet{proposition:ebcf409125ae6be5-5}{%
\ \ \ \ \isacommand{from}\isamarkupfalse%
\ {\isacartoucheopen}p\ {\isasymlongleftrightarrow}\ q{\isacartoucheclose}\ \isacommand{show}\isamarkupfalse%
\ {\isacartoucheopen}q\ {\isasymLongrightarrow}\ p{\isacartoucheclose}\ \isacommand{{\isachardot}{\kern0pt}{\isachardot}{\kern0pt}}\isamarkupfalse%
\isanewline
}
\DefineSnippet{proposition:ebcf409125ae6be5-6}{%
\ \ \isacommand{next}\isamarkupfalse%
\isanewline
}
\DefineSnippet{proposition:ebcf409125ae6be5-7}{%
\ \ \ \ \isacommand{from}\isamarkupfalse%
\ {\isacartoucheopen}p\ {\isasymlongleftrightarrow}\ q{\isacartoucheclose}\ \isacommand{show}\isamarkupfalse%
\ {\isacartoucheopen}p\ {\isasymLongrightarrow}\ q{\isacartoucheclose}\ \isacommand{{\isachardot}{\kern0pt}{\isachardot}{\kern0pt}}\isamarkupfalse%
\isanewline
}
\DefineSnippet{proposition:ebcf409125ae6be5-8}{%
\ \ \isacommand{qed}\isamarkupfalse%
\isanewline
}
\DefineSnippet{proposition:ebcf409125ae6be5-9}{%
\isacommand{next}\isamarkupfalse%
\isanewline
}
\DefineSnippet{proposition:ebcf409125ae6be5-10}{%
\ \ \isacommand{assume}\isamarkupfalse%
\ {\isacartoucheopen}q\ {\isasymlongleftrightarrow}\ p{\isacartoucheclose}\isanewline
}
\DefineSnippet{proposition:ebcf409125ae6be5-11}{%
\ \ \isacommand{show}\isamarkupfalse%
\ {\isacartoucheopen}p\ {\isasymlongleftrightarrow}\ q{\isacartoucheclose}\isanewline
}
\DefineSnippet{proposition:ebcf409125ae6be5-12}{%
\ \ \isacommand{proof}\isamarkupfalse%
\isanewline
}
\DefineSnippet{proposition:ebcf409125ae6be5-13}{%
\ \ \ \ \isacommand{from}\isamarkupfalse%
\ {\isacartoucheopen}q\ {\isasymlongleftrightarrow}\ p{\isacartoucheclose}\ \isacommand{show}\isamarkupfalse%
\ {\isacartoucheopen}p\ {\isasymLongrightarrow}\ q{\isacartoucheclose}\ \isacommand{{\isachardot}{\kern0pt}{\isachardot}{\kern0pt}}\isamarkupfalse%
\isanewline
}
\DefineSnippet{proposition:ebcf409125ae6be5-14}{%
\ \ \isacommand{next}\isamarkupfalse%
\isanewline
}
\DefineSnippet{proposition:ebcf409125ae6be5-15}{%
\ \ \ \ \isacommand{from}\isamarkupfalse%
\ {\isacartoucheopen}q\ {\isasymlongleftrightarrow}\ p{\isacartoucheclose}\ \isacommand{show}\isamarkupfalse%
\ {\isacartoucheopen}q\ {\isasymLongrightarrow}\ p{\isacartoucheclose}\ \isacommand{{\isachardot}{\kern0pt}{\isachardot}{\kern0pt}}\isamarkupfalse%
\isanewline
}
\DefineSnippet{proposition:ebcf409125ae6be5-16}{%
\ \ \isacommand{qed}\isamarkupfalse%
\isanewline
}
\DefineSnippet{proposition:ebcf409125ae6be5-17}{%
\isacommand{qed}\isamarkupfalse%
\endisatagproof
{\isafoldproof}%
\isadelimproof
\isanewline
}
\DefineSnippet{proposition:ebcf409125ae6be5-18}{%
\endisadelimproof
\isadelimtheory
}

\newenvironment{qisabelle}{\begin{quote}\begin{isabelle}}{\end{isabelle}\end{quote}}

\newcommand{\Snippet}[1]{%
  \newcount\i
  \i=0
  \loop
    \csname snippet--#1-\the\i\endcsname
    \advance \i 1
  \ifcsname snippet--#1-\the\i\endcsname
  \repeat
}

\newcommand{\SnippetPart}[3]{%
{%
  \newcount\i
  \i=#1
  \loop
    \ifnum \i=#2
      \renewcommand{\isanewline}{}%
    \fi
    \csname snippet--#3-\the\i\endcsname
    \advance \i 1
    \ifnum \i>#2 {}
    \else \repeat
}
}

\def\isadelimtheory{}\def\endisadelimtheory{}
\def\isatagtheory{}\def\endisatagtheory{}

\def\isadelimproof{}\def\endisadelimproof{}
\def\isatagproof{}\def\endisatagproof{}
\def\isafoldproof{}

\def\isacartoucheopen{\isatext{\raise.3ex\hbox{$\scriptscriptstyle\langle\,\,\,$}}}%
\def\isacartoucheclose{\isatext{\raise.3ex\hbox{$\scriptscriptstyle\,\,\,\rangle$}}}%

\title{Teaching Intuitionistic and Classical Propositional Logic Using Isabelle}

\def\titlerunning{Teaching Intuitionistic and Classical Propositional Logic Using Isabelle}

\author{Jørgen Villadsen \qquad Asta Halkjær From 
\institute{DTU Compute - Department of Applied Mathematics and Computer Science - Technical University of Denmark}
\and
Patrick Blackburn
\institute{Section of Philosophy and Science Studies, IKH, Roskilde University, Denmark}}

\def\authorrunning{J. Villadsen, A. H. From \&\ P. Blackburn}

\begin{document}

\maketitle

\begin{abstract}
We describe a natural deduction formalization of intuitionistic and classical propositional logic in the Isabelle/Pure framework. In contrast to earlier work, where we explored the pedagogical benefits of using a deep embedding approach to logical modelling, here we employ a logical framework modelling. This gives rise to simple and natural teaching examples and we report on the role it played in teaching our Automated Reasoning course in 2020 and 2021.
\end{abstract}

\section{Introduction}
\label{intro}

We describe a formalization of intuitionistic and classical propositional logic in the Isabelle/Pure framework~\cite{Pure08,Isar08,Paulson89}.
In a few lines of code we axiomatize our choice of connectives with their natural deduction introduction and elimination rules. 
This gives meaning to the (object) logic via the (meta) Isabelle/Pure framework rather than in potentially ambiguous natural language.
Unlike pen-and-paper proofs that must be manually checked for mistakes, Isabelle can check automatically, reducing the burden on lecturers and teaching assistants.
As we will see, by decorating the rules with appropriate attributes, we never have to mention exactly which rule is being applied, as the context determines it.
This results in proofs that are natural to write and natural to read. 

Our work is based on the Pure/Examples by Makarius in the Isabelle sources:
\begin{center}
    \url{https://isabelle.in.tum.de/dist/library/Pure/Pure-Examples/document.pdf}
\end{center}
That document, however, contains few comments and we have polished the formalization and tested it in class.
Our formalization is available online in various formats:
\begin{center}
    \url{https://hol.compute.dtu.dk/Pure_I/}
\end{center}
We have used our approach in our Automated Reasoning course in 2020 and 2021:
\begin{center}
    \url{https://kurser.dtu.dk/course/02256}
\end{center}
This is an advanced MSc course (about 40 students) with a focus on the natural deduction proof system, first-order logic, higher-order logic and type theory, in particular Isabelle/HOL~\cite{nipkow+02}.

We call the approach Pure-I (“pure first”) for two reasons:
\begin{itemize}
\item 
We use Isabelle/Pure as a stepping stone towards the default higher-order logic in Isabelle/HOL
\item
We can understand the letter "I" both as a reference to the use of Isabelle by the students (the tools are a series of Isabelle theory files) but also for Intuitionism since we let the students add the Law of Excluded Middle (LEM) as the axiom \( p \lor \neg p \) when needed for classical reasoning.
\end{itemize}
We note that Isabelle/HOL is built on classical logic but with a built-in proof method for intuitionistic logic (\emph{iprover}).

We focus on propositional logic but extensions to first-order logic as well as higher-order logic are possible.
We use the same natural deduction rule names as in our natural deduction assistant NaDeA~\cite{ThEdu17}.
We teach NaDeA first but this is not a requirement.
However, by teaching NaDeA first we can move quicker through the material as the rules are already familiar to the students and only the Isabelle setting is new.
This would also be possible if the students had already learned natural deduction in some other way.

Ideally, our students participating in the MSc course have already taken our advanced BSc course on Logical Systems and Logic Programming (about 80 students):
\begin{center}
    \url{https://kurser.dtu.dk/course/02156}
\end{center}
In practice this is not the case, in particular due to a large intake of international students, so we start the MSc course with some weeks on axiomatic systems and/or sequent calculus presented in a different way than in the BSc course \cite{FOL-Axiomatic-AFP,PAAR}.

\begin{figure}[t]
\centering
\includegraphics[width=0.80\textwidth]{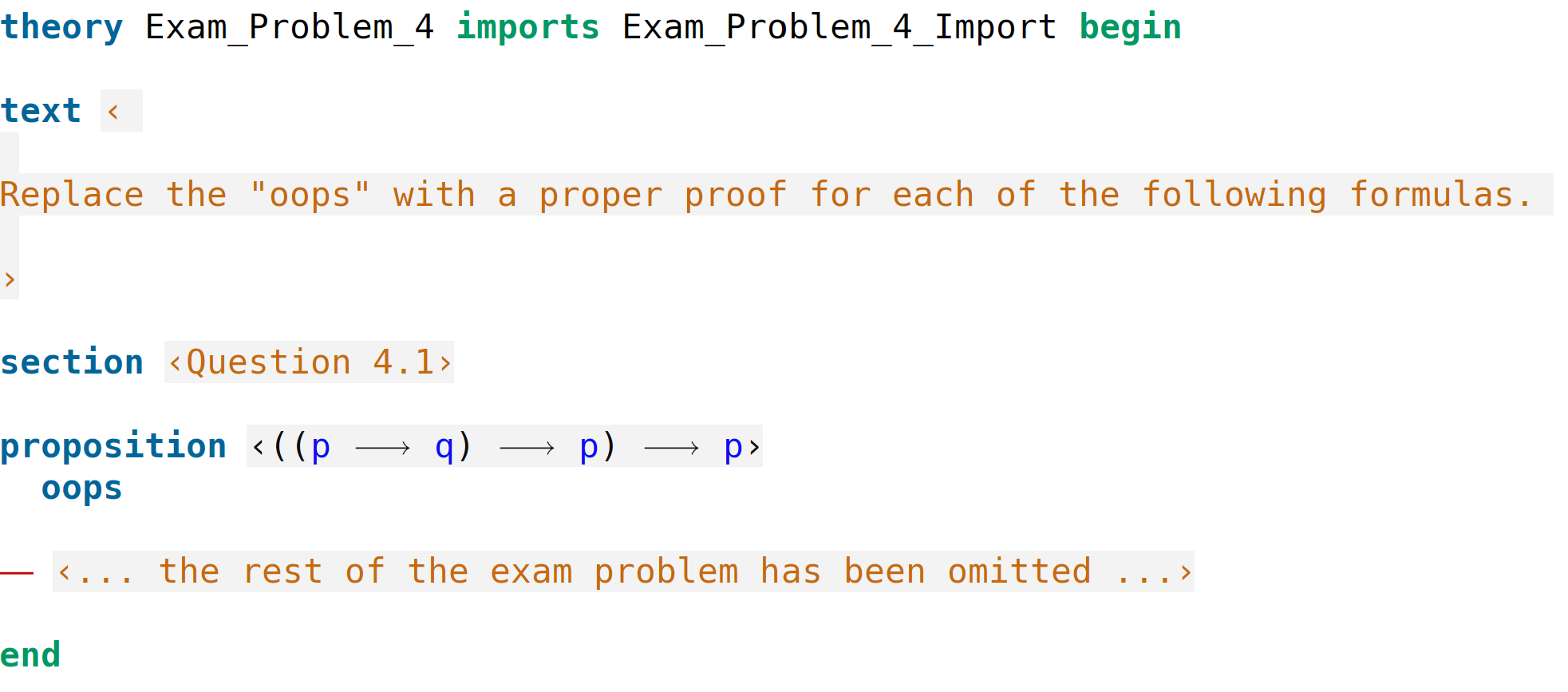}
\caption{Part of an exam question using our formalization.}%
\label{fig:Exam}
\end{figure}

Figure~\ref{fig:Exam} shows an actual question for the written exam, 27 May 2021, in our course 02256 Automated Reasoning.
The solution has since been added as theorem \isa{Peirce} in Fig.~\ref{fig:Classical}.

The exam was 2 hours with all aids.
36 students registered and 30 students passed.

The distribution of the grades is available online:
\begin{center}
    \url{https://karakterer.dtu.dk/Histogram/1/02256/Summer-2021}
\end{center}
In Denmark we use a so-called 7-step-scale, designed to be compatible with the ECTS grading scale, with A=12, B=10, C=7, D=4, E=02, Fx=00 and F=-3.
Two students were not graded because they did not show up at the exam.
The grade average were between B and C.
This is quite a high grade average but not unusual for advanced courses.

In the next section we provide an overview of our teaching and the present paper.

\newpage

\section{Overview of Teaching and Paper}

\begin{table}[t]
\caption{Topics in our course along with relevant tasks (parentheses denote optional tasks).}%
\label{tab:overview}
\bigskip
\begin{tabular}{l l l l}
    Topic & Section & Figures & Tasks \\[1ex]
    \midrule \\[0ex]
    Logical Frameworks & \ref{log-frame} & & Study the Isabelle/Pure framework \\[1ex]
    Axiomatizations & \ref{axioms} & & Study the Pure-I formalization \\[1ex]
    Definitions & \ref{definitions} & & Study the derived rules of defined connectives \\[1ex]
    Examples & \ref{simple-example} & \ref{fig:Pure-I-Verbose} &
          Study the Figure \\[1ex]
    & & & Do at least a dozen proofs in intuitionistic logic \\[1ex]
    & & & (Port to Isabelle/HOL: replace Pure-I with Main\footnotemark) \\[1ex]
    Classical Logic & \ref{classical} & \ref{fig:Classical}, \ref{fig:ClassicalVerbose} & Study the first Figures \\[1ex]
    & & & Do a few proofs in classical logic \\[1ex]
    & & (\ref{fig:Classical-NaDeA}, \ref{fig:Classical-NaDeA-Appendix}, \ref{fig:Main-HOL}, \ref{fig:Main-HOL-Appendix}) & (Do the proofs in the remaining Figures) \\[1ex]
    Exam preparation & - & \ref{fig:Exam} & Do at least a handful of proofs. \\[5ex]
\end{tabular}
\end{table}

\footnotetext{Main is the Isabelle theory that defines Isabelle/HOL whereas Pure-I is our axiomatization.}

We explain the following approaches in the following Section~\ref{log-frame}:
\begin{itemize}
    \item Deep Embedding
    \item Shallow Embedding
    \item Logical Framework
\end{itemize}

Our use of Pure-I in the course proceeds in the following way.
We ask students to install Isabelle before the course starts:
\begin{center}
\url{https://isabelle.in.tum.de/}
\end{center}

As preparation, they can download and open Pure-I (see above) or if Isabelle is not installed, browse the formalization here:
\begin{center}
\url{https://hol.compute.dtu.dk/Pure_I/Pure_I.html}
\end{center}

We then cover the topics in Table~\ref{tab:overview}, which are described further in the listed sections and figures.
We have also included tasks for the students that are relevant to understanding the material.

In Section~\ref{ped-ben} we take a step back and consider the pedagogical benefits of this approach from a more general perspective, before we conclude in Section~\ref{conc}.

\newpage

\section{Logical Frameworks vs. Deep/Shallow Embeddings}
\label{log-frame}

In this paper we present a simple and direct way of teaching propositional logic, namely by ``axiomatizing'' the Isabelle/Pure system in the most direct way possible.
The pedagogical benefits of this approach are summed up in the words `simple' and `natural'. The object and metalevel deductions mirror each other clearly --- and the way that teachers would typically instruct students to construct proofs is directly reflected in the code the students learn to write to prove things in Isabelle (cf.~Section~\ref{simple-example}). 

We have used this system as part of a course on automated reasoning taught at the Technical University of Denmark (DTU); the course aimed to teach both natural deduction and Isabelle.
It is the latest in a line of systems (based on courses taught at DTU) which explores different ways in which Isabelle can be used for pedagogical purposes~\cite{ThEdu20,ThEdu19,ThEdu18,ThEdu17}.
For example, in the paper presented at the 2020 edition of ThEdu, we described a formalization of a simple axiomatization, called W~\cite{ThEdu20}, for classical propositional logic.
This formalization was built using a \emph{deep embedding} of System W in the higher-order logic of Isabelle/HOL:

\begin{trivlist}
\item \textbf{Deep Embedding} \hfill
\begin{minipage}[t]{0.75\textwidth}
The act of representing one language, typically a logic or programming language, with another by modeling expressions in the former as data in the latter.
\end{minipage}
\medskip

\hfill-- Wiktionary, \url{https://en.wiktionary.org/wiki/deep_embedding}
\end{trivlist}

Deep embedding is certainly more appropriate when it comes to teaching many aspects of metatheory --- it forces the student to get to grips with what is involved in formalizing a language \emph{and its semantics}, and working in this way opens a natural path to studying such topics as soundness, completeness and interpolation (and learning how to use Isabelle/HOL to prove such results).
This has certain pedagogical advantages --- we argued in the paper that this approach forces students to consider metatheoretic issues right from the start.
Deep embeddings are good for this purpose: for example, they make it possible to teach structural induction on the syntax (which is a key metalogical proof tool) as the object language syntax has a concrete representation in the metalogic.

Here, however, we want to take a step back from deep embedding.
We want a gentler approach better suited for a first course in logic, especially for students with less technical backgrounds.
Part of the goal is to teach \textit{logic} (rather than metalogic) and in particular to teach natural deduction.
The simple logical framework approach used in this paper enables us to directly ``cook up'' our Isabelle/Pure formalization from scratch without worrying about higher-order logic but while still teaching skills that are transferable to Isabelle/HOL.

We see the chief benefits of this lighter approach as being simplicity and elegance.
This simplicity and elegance traces to two sources: the fact that we are working with natural deduction and the ease with which such proof rules can be reflected in Isabelle/Pure.

We start by introducing intuitionistic propositional logic and only afterwards axiomatize the Law of Excluded Middle (LEM).
Intuitionistic propositional logic is simple in a number of ways. For example, every statement of the form $\varphi \lor \psi$ is effectively transformable into an intuitionistic proof of $\varphi$ or an intuitionistic proof of $\psi$.
Moreover, intuitionistic propositional logic is effectively decidable: a finite constructive process applies uniformly to every formula, and this process will either produce an intuitionistic proof of the formula or demonstrate that no such proof can exist.
When working in the intuitionistic fragment, our axiomatization embodies this process.
In other words: it directly reflects, in very simple code, the process of natural deduction proofs.

We then add some complexity by moving to classical propositional logic.
Working with intuitionistic proofs has built familiarity with the system, preparing the student to handle the more complicated classical proofs where there is no longer a clear recipe for how to prove a formula.
As we will see, however, the move to classical logic does not require any new techniques and can therefore easily be explained.

Besides the logical framework approach taken here and the deep embeddings considered previously~\cite{ThEdu20}, there remains a third option to explore:

\begin{trivlist}
\item \textbf{Shallow Embedding} \hfill
\begin{minipage}[t]{0.75\textwidth}
The act of representing one logic or language with another by providing a syntactic translation.
\end{minipage}
\medskip

\hfill-- Wiktionary, \url{https://en.wiktionary.org/wiki/shallow_embedding}
\end{trivlist}

Shallow embeddings are another approach to formalizing logics, one that is widely used in the Isabelle community.
Moreover, it is an important way of thinking about many topics in logic in its own right. Modal logic, for example, can be thought of in terms of Kripke models (a deep embedding perspective) but some logicians prefer to emphasize the standard translation of modal logic into first-order logic (a shallow embedding perspective).
It would be interesting to use a shallow embedding of modal logic in Isabelle/HOL as a teaching tool that emphasised the standard translation perspective on modal logic.
In terms of modal logic and Isabelle/Pure, Dion\'{i}sio et al.~\cite{DionisioGM05} have successfully used Isabelle/Pure to teach hybrid logic to students in an approach similar to ours.
The students even define their own labelled modal logic~\cite{BasinMV98} in the end.

To conclude, in both the deep and shallow embedding approaches, we are forced to grapple with the complexity of Isabelle/HOL as our metalogic.
Here, we only rely on the simple logical framework of Isabelle/Pure.
This leaves more room for our goal: teaching logic using Isabelle.

\section{Axiomatizations - Pure-I}
\label{axioms}

In our approach we start at the bottom by declaring the type of formulas in our object logic, namely boolean:

\begin{qisabelle}
    \Snippet{typedecl:f7b94bb19014cba6}
\end{qisabelle}

We need to link this type with Isabelle's type of propositions using a so-called ``truth judgment.''
This automatically lifts our formulas so we can use the machinery provided by the Isabelle/Pure framework.
The declaration is simple:

\begin{qisabelle}
    \Snippet{judgment:Trueprop}
\end{qisabelle}

We are now ready to axiomatize our first connective:

\begin{qisabelle}
    \Snippet{axiomatization:Imp}
    \end{qisabelle}
    
The first line declares a name, \isa{Imp}, for our connective and an infix symbol, \isa{\isasymlongrightarrow}, with right associativity and precedence 3.
The next two lines describe the meaning via an introduction and an elimination rule.
Here we see the power of using the Isabelle/Pure framework.
Instead of using annotations, boxes or similar to discharge assumptions, we simply defer to the meta-logic's implication, \isa{\isasymLongrightarrow}, to explain our object-logic's implication.
That is, to prove that \isa{p} implies \isa{q}, assume \isa{p} at the meta level and prove \isa{q}, letting the proof assistant handle the details.
The elimination rule, modus ponens, is symmetric and lifts the object-level implication into the meta-level, letting us prove \isa{q} from \isa{p} when we know \isa{p {\isasymlongrightarrow} q}.

We consider the axiomatization of disjunction next:

\begin{qisabelle}
    \Snippet{axiomatization:Dis}
\end{qisabelle}

Again we declare a name, \isa{Dis}, and an infix symbol, \isa{\isasymor}.
This time we have one elimination rule and two introduction rules.
The elimination rule allows us to conclude \isa{r} from \isa{p {\isasymor} q} if we can prove \isa{r} from both \isa{p} and \isa{q}.
Each of \isa{p}, \isa{q} and \isa{r} are meta-variables that can be instantiated for any formula.
The two introduction rules allow us to form a disjunction when we have proved either side of it.
The attributes on the rules, \isa{intro} and \isa{elim}, will allow us to use them without explicit reference in the proofs later on.

The axiomatization of conjunction uses the same principles as we have already seen:

\begin{qisabelle}
    \Snippet{axiomatization:Con}
\end{qisabelle}

Finally we axiomatize falsity, \isa{\isasymbottom}, with just an elimination rule, the principle of explosion:

\begin{qisabelle}
    \Snippet{axiomatization:Falsity}
\end{qisabelle}

Experienced Isabelle users will see that we have used it in a particularly simple and direct way: we have turned Isabelle into an ``instance'' of intuitionistic logic by adding axioms.

\section{Definitions - Pure-I}
\label{definitions}

Besides axiomatizations we can provide simple definitions that hide an expression under a name.
In this way, we can define the dual of \isa{\isasymbottom}, a symbol, \isa{\isasymtop}, that denotes an always true formula:

\begin{qisabelle}
    \Snippet{definition:Truth}
\end{qisabelle}

We can still give a meaning to this, as an introduction rule, so we do not need to consider the implementation details later.
Instead of being part of an axiomatization, this introduction rule is stated simply as a theorem:

\begin{qisabelle}
    \Snippet{theorem:Truth-I}
\end{qisabelle}

Several concepts are at play here.
In the first line we state the name of the theorem, the attributes and the goal \isa{\isasymtop}.
Next we unfold the definition, transforming the goal into \isa{{\isasymbottom} {\isasymlongrightarrow} {\isasymbottom}}.
The two periods \isacommand{..} apply a matching introduction rule for the connective, in this case \isa{Imp-I}.
That reduces the goal to \isa{{\isasymbottom} {\isasymLongrightarrow} {\isasymbottom}}, which is discharged automatically.

We define negation in the usual way, as implication of a contradiction:

\begin{qisabelle}
    \Snippet{definition:Neg}
\end{qisabelle}

This brings us to a more interesting theorem and proof: from a formula and its negation we can prove anything.
We start by stating the theorem and unfolding the definition:

\begin{qisabelle}
    \Snippet{theorem:Neg-E}
\end{qisabelle}

The \isacommand{proof} command starts the proof and the following hyphen tells Isabelle not to apply any initial proof rules to transform the goal (further).
The \isacommand{assume} command on the next line is implication introduction at the meta-level: it introduces the two \isa{\isasymLongrightarrow} connectives in the goal.
Since we unfolded the definition of negation, the first assumption is \isa{p {\isasymlongrightarrow} {\isasymbottom}}, while the second one is \isa{p} as expected.
The \isacommand{then} command makes these two assumptions (or any other just stated fact) available to the following command, which in this case is \isacommand{have}.
This command is used to state local results and here we use it to prove \isa{\isasymbottom} from the two assumptions.
The two periods~\isacommand{..} automatically apply the \isa{Imp-E} rule to \isa{p~{\isasymlongrightarrow}~{\isasymbottom}} and \isa{p} in the proof context.
The \isacommand{show} on the penultimate line states a result like \isacommand{have} but also discharges a goal, in our case \isa{q}.
Here the two periods automatically apply the \isa{Falsity-E} rule.
The final \isacommand{qed} matches the initial \isacommand{proof} and marks the end of the proof.
It only ``succeeds'' if there are no remaining goals.

We define bi-implication as implication in both directions:

\begin{qisabelle}
    \Snippet{definition:Iff}
\end{qisabelle}

Again, we want to prove introduction and elimination rules that allow us to work with the connective without unfolding its definition.
We start with the introduction rule.
Given that we can prove \isa{q} from \isa{p} and \isa{p} from \isa{q} we seek to prove \isa{p {\isasymlongleftrightarrow} q}:

\begin{qisabelle}
    \Snippet{theorem:Iff-I}
\end{qisabelle}

We have already seen the \isacommand{proof} and \isacommand{assume} commands, which are used in the same way here.
Instead of using \isacommand{then}, we pick out the assumptions one at a time with \isacommand{from} in the next two lines, introducing the object-logic implication in each direction.
The final line combines the two implications into their conjunction: the definition of bi-implication.

The proofs of the two elimination rules only use techniques we have already seen, but some other axiomatized rules, making them suitable as exercises:

\begin{qisabelle}
    \Snippet{theorem:Iff-E1}
\end{qisabelle}

\begin{qisabelle}
    \Snippet{theorem:Iff-E2}
\end{qisabelle}

\section{Selected Teaching Examples}
\label{simple-example}

Let us examine some examples.
The \isacommand{proposition} command is an alternative name for \isacommand{theorem} and we use it for student exercises and examples.
There is no difference to the system, only to the human reader.
\begin{qisabelle}
    \Snippet{proposition:c034bbb10ac00aeb}
\end{qisabelle}
Note the way the Isabelle/Pure code matches the way we would explain to students how they should go about proving this in natural deduction. It essentially spells out a constructive process for building the proof tree required --- with the formula being proved if and only if such a proof tree exists.

\begin{figure}[t]
\centering
\includegraphics[width=0.7\textwidth]{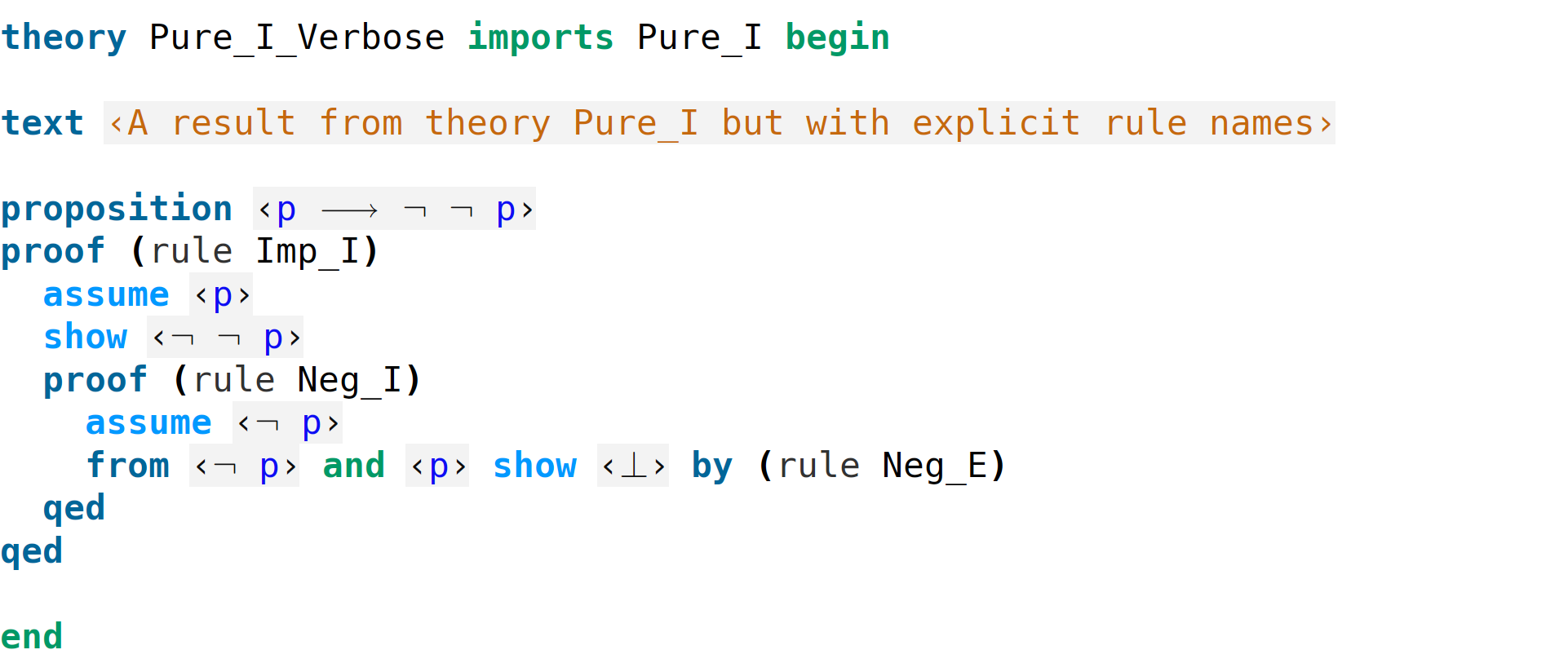}
\caption{A simple teaching example with explicit rule names in Isabelle/Pure.}%
\label{fig:Pure-I-Verbose}
\end{figure}

In Fig.~\ref{fig:Pure-I-Verbose} we see the same example but with explicit rule names like \isa{Imp\_I} and \isa{Neg\_I} (and syntax highlighting as seen by the students).
It is always possible to write explicitly which rule justifies a proof step, either for clarity or to convince oneself that the implicitly applied rule is the expected one.
In general, however, we prefer to leave such rules implicit, so the structure of the proof stands out clearly.

Here is a proof of modus tollens. Once again, note how Isabelle/Pure tracks the way a teacher would explain to a beginner how to construct the relevant proof tree:

\begin{qisabelle}
    \Snippet{proposition:88d0827f917d9c91}
\end{qisabelle}

Here is a more complex example:

\begin{qisabelle}
    \Snippet{proposition:ebcf409125ae6be5}
\end{qisabelle}

Here, we need to prove both directions of the middle bi-implication by assuming the left-hand side and showing the right and vice versa.
Conceptually, the \isacommand{next} command switches to proving the remaining direction.
Technically, it resets the local context, so we cannot use the assumption from the proof in one direction in the proof of the other direction.
With this command at hand, the proof almost writes itself.

The above example works verbatim in Isabelle/HOL.
This makes the skills obtained by working with this logic directly transferable to the larger setting of higher-order logic.
The syntax is the same and the same rules are available, making this a stepping stone for proving things in Isabelle/HOL.

\section{Classical Logic}
\label{classical}

Imagine we try to prove a formula that does not have a proof in our logic, e.g.~Peirce's law:

\[ ((p \longrightarrow q) \longrightarrow p) \longrightarrow p \]

We can take a stab at it, applying introduction rules for the implications and so on, but we will never succeed.
Like with pen and paper, the system does not provide any indication that no proof exists, but unlike with pen and paper we cannot mistakenly think one does by writing down a flawed proof.
That specific formula does hold in classical propositional logic and so we could extend the axiomatization with the Law of Excluded Middle (LEM) \( p \lor \neg p \) to work in that logic instead.
Then we would be able to find a proof.

\begin{figure}[t]
\centering
\includegraphics[width=0.7\textwidth]{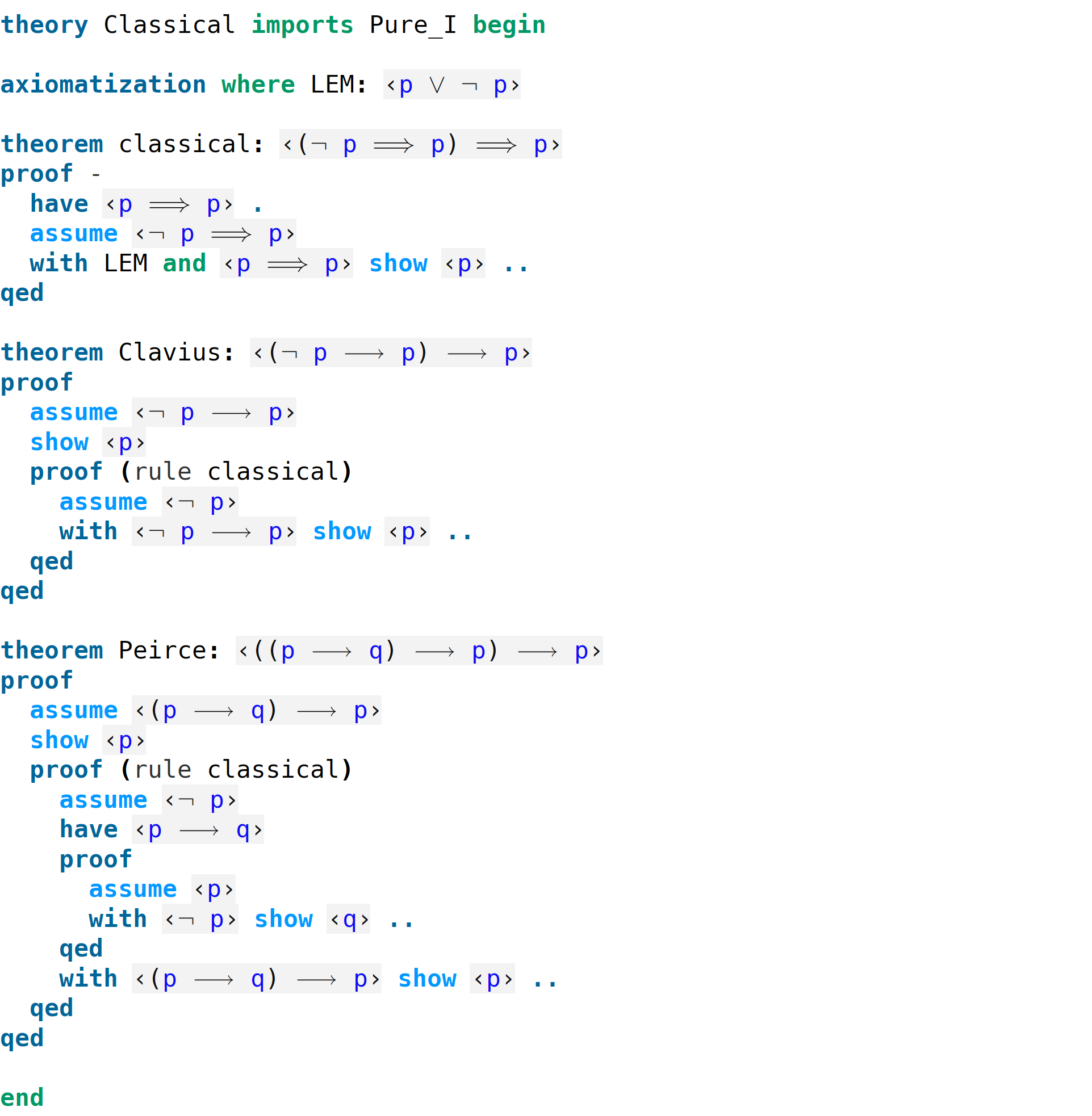}
\caption{Axiomatization of the Law of Excluded Middle (LEM) with subsequent proofs of three other principles in classical logic.}%
\label{fig:Classical}
\end{figure}

In Fig.~\ref{fig:Classical} we do so by axiomatizing the Law of Excluded Middle (LEM) on top of Pure-I.
This axiomatization of a simple formula is easy to explain to students and clearly states an important point in classical logic: for every proposition, either this proposition or its negation is true.
We first derive a more useful principle \isa{classical}, named after its counterpart in Isabelle/HOL.
The single period command proves immediate facts like \isa{p \isasymLongrightarrow\ p}.
The \isacommand{with} command makes one or more assumptions or previously established facts available as justification for the following statement.
We use the theorem \isa{classical} to prove Clavius's law and Peirce's law; we can no longer rely only on two periods as for intuionistic proofs, but must explicitly invoke a classical principle in the proof.

\begin{figure}[t]
\centering
\includegraphics[width=0.7\textwidth]{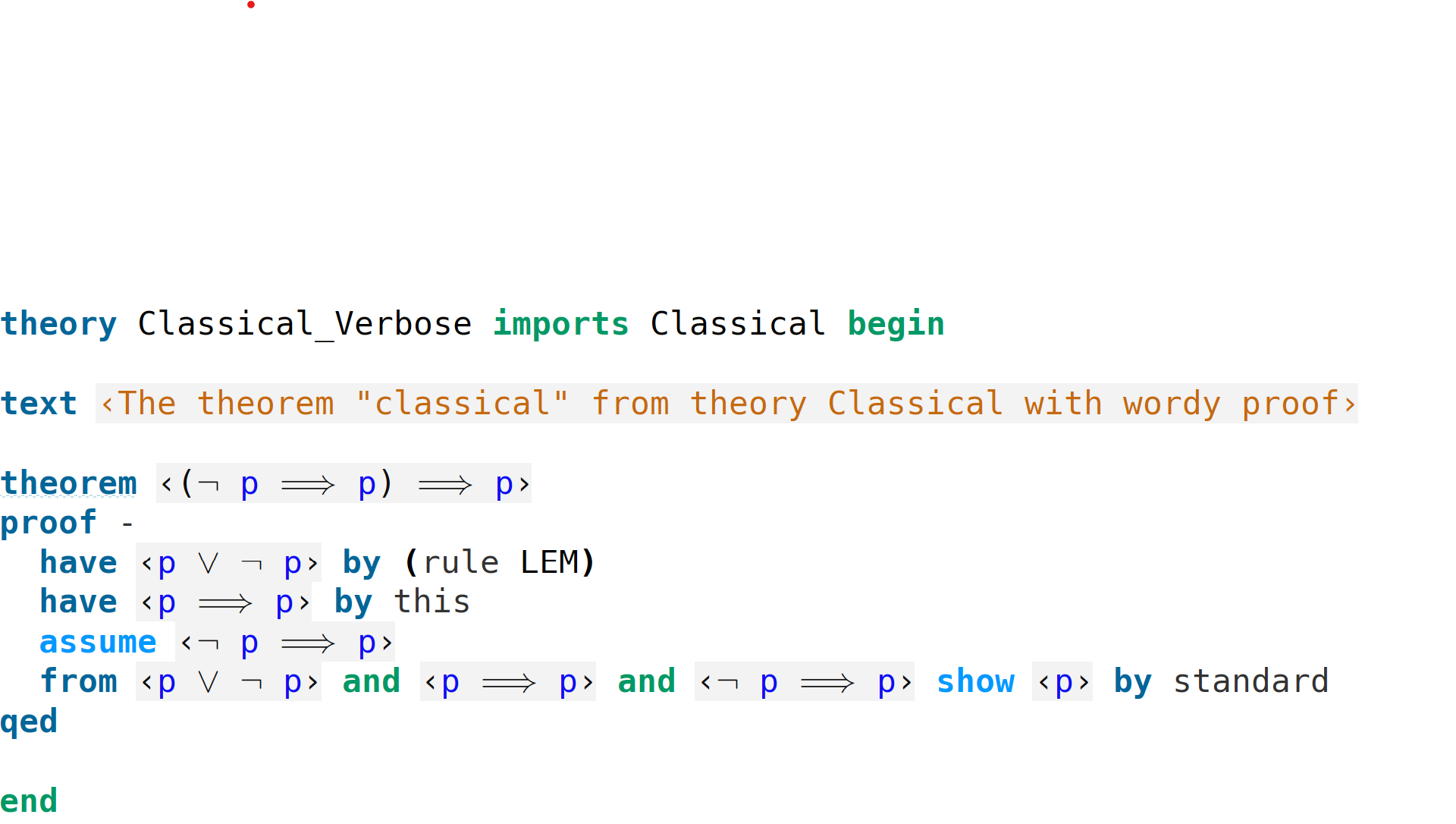}
\caption{A proof from Fig.~\ref{fig:Classical} with explicit rules.}%
\label{fig:ClassicalVerbose}
\end{figure}

In Fig.~\ref{fig:ClassicalVerbose} we give a wordy proof of the \isa{classical} theorem from Fig.~\ref{fig:Classical}.
Here we use the full name \isa{this} for the single period proof method and \isa{standard} for the double period.
We also state \isa{LEM} directly and use \isacommand{from} instead of \isacommand{with} to be more explicit.
It is often useful for students to work with variants of the key proofs. 

\begin{figure}[t]
\centering
\includegraphics[width=0.7\textwidth]{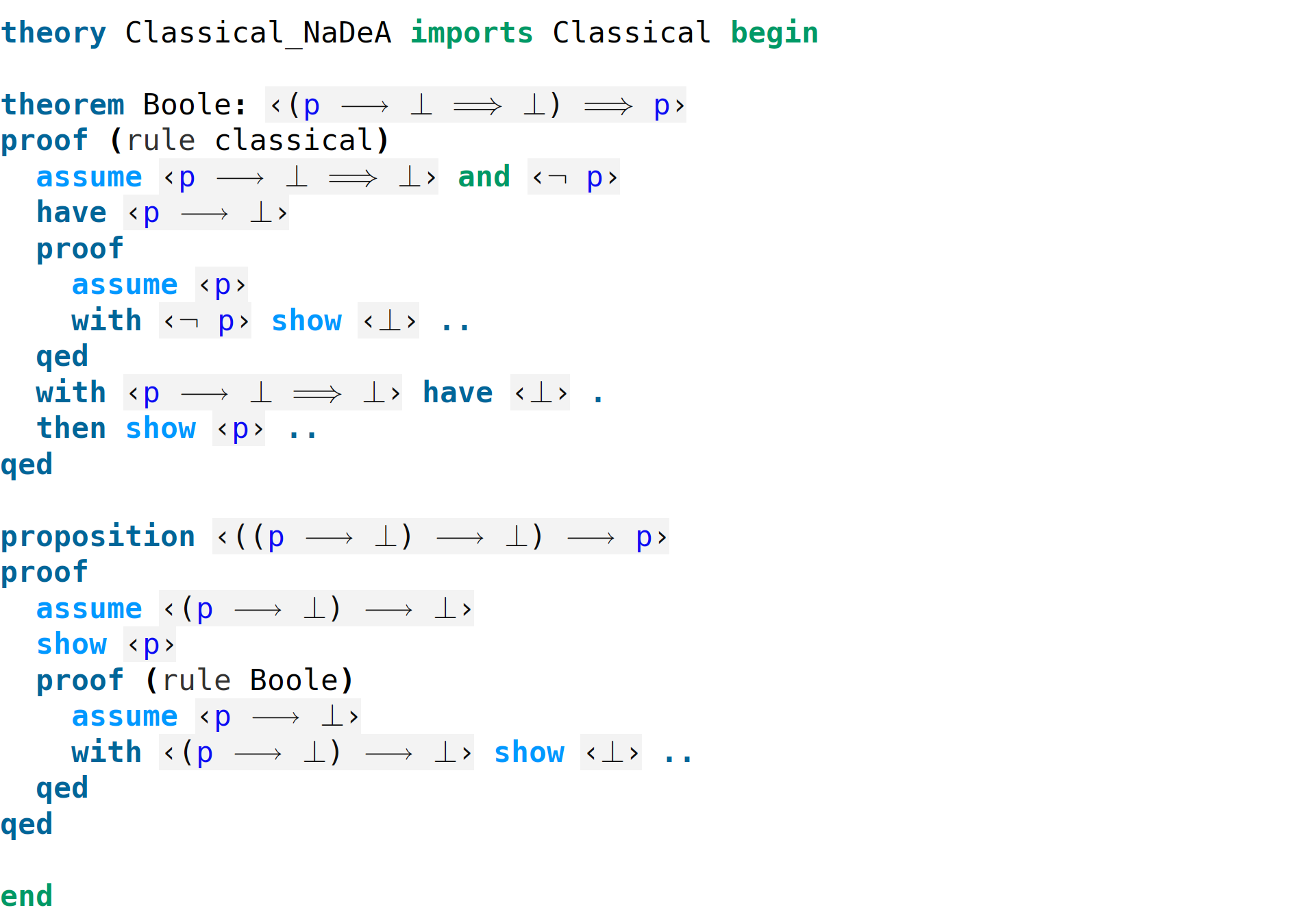}
\caption{A proof of the classical principle Boole used in NaDeA and a subsequent proof.}%
\label{fig:Classical-NaDeA}
\end{figure}

In Fig.~\ref{fig:Classical-NaDeA} we consider the special NaDeA rule Boole in the Isabelle/Pure setting.
After proving it from \isa{classical} (and transitively \isa{LEM}), we prove a variant of double negation elimination using it.
Formulas like this can be used either for demonstration or as student exercises.

\begin{figure}[t]
\centering
\includegraphics[width=0.7\textwidth]{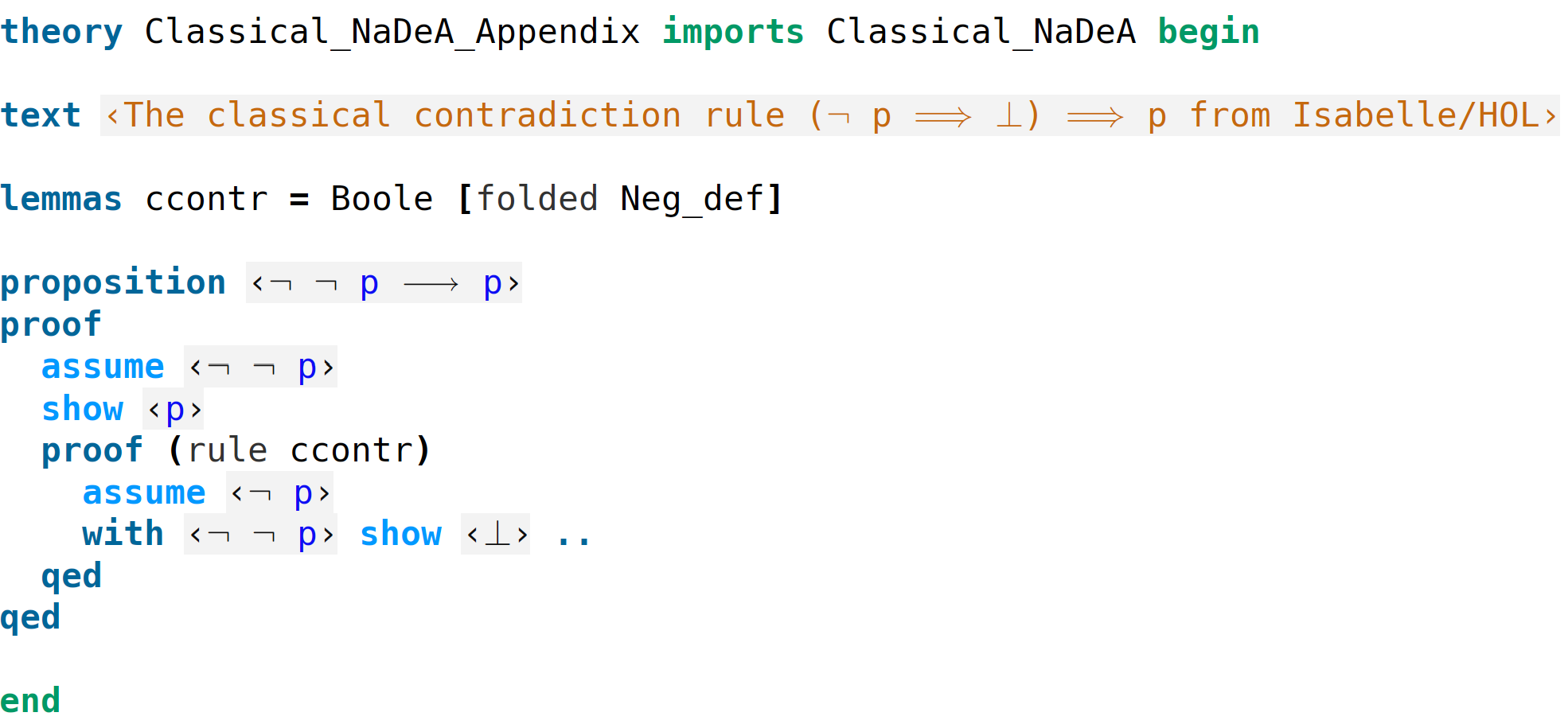}
\caption{A classical proof using the principle Boole from NaDeA formulated as in Isabelle/HOL.}%
\label{fig:Classical-NaDeA-Appendix}
\end{figure}

In Fig.~\ref{fig:Classical-NaDeA-Appendix} we consider classical contradiction (\isa{ccontr} in Isabelle/HOL): if from \isa{\isasymnot\ p} we can derive falsity, then we can conclude \isa{p}.
We have already proved this as Boole when we recall that we have defined \( \neg p \equiv p \longrightarrow \bot \).
We then prove the same formula as in Fig.~\ref{fig:Classical-NaDeA} but stated using negation.

\begin{figure}[t]
\centering
\includegraphics[width=0.75\textwidth]{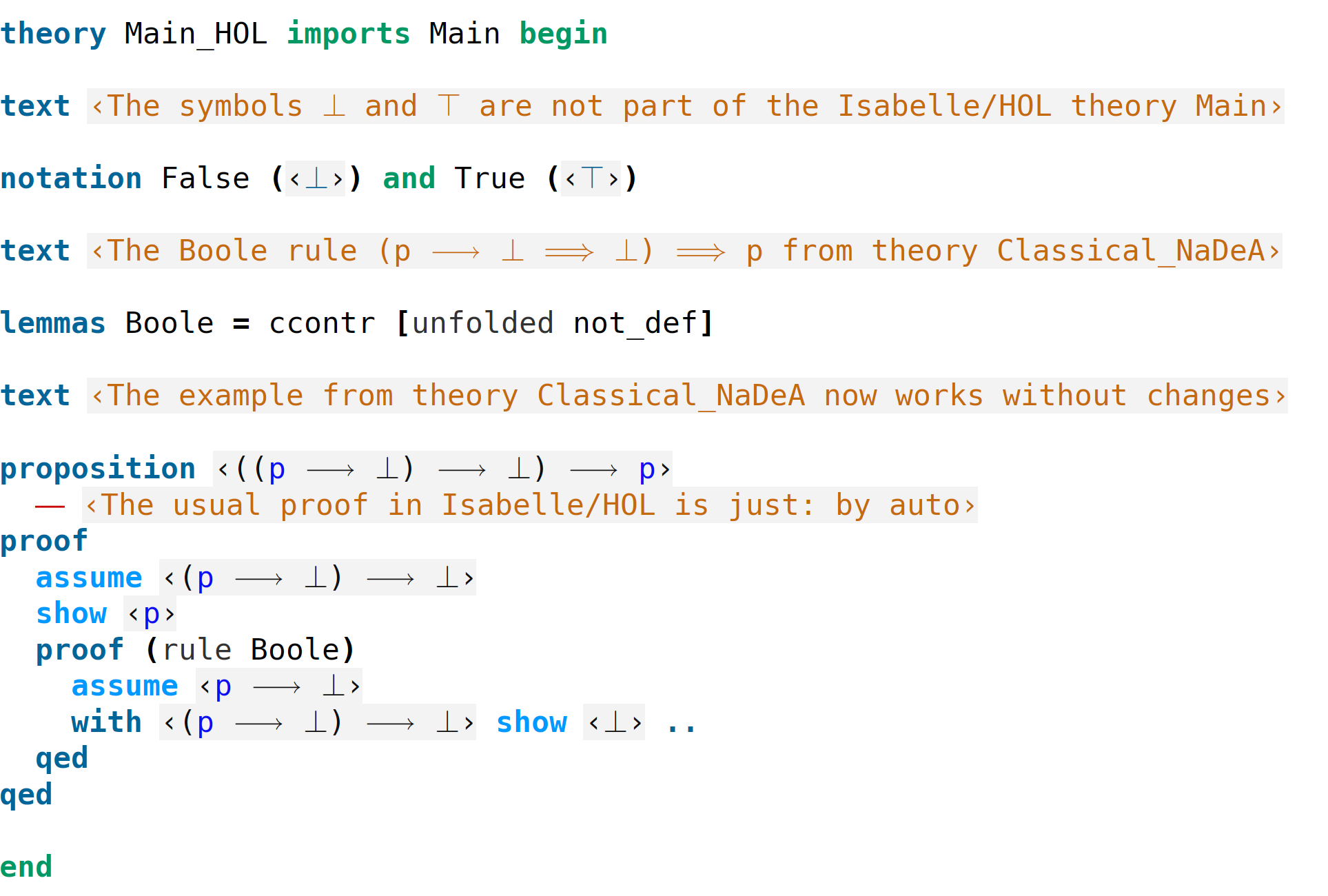}
\caption{A proof from Fig.~\ref{fig:Classical-NaDeA} after a few definitions in Isabelle/HOL.}%
\label{fig:Main-HOL}
\end{figure}

In Fig.~\ref{fig:Main-HOL} we define \isa{\isasymbottom} and \isa{\isasymtop} in Isabelle/HOL as \isa{False} and \isa{True}, respectively.
We then define the Boole rule from the pre-defined \isa{ccontr} by unfolding negation similarly to previously.
Then we can insert our proof from Fig.~\ref{fig:Classical-NaDeA} unmodified.
This demonstrates how the skills obtained from working with our formalization in Isabelle/Pure are transferable to the full Isabelle/HOL.

\begin{figure}[t]
\centering
\includegraphics[width=0.75\textwidth]{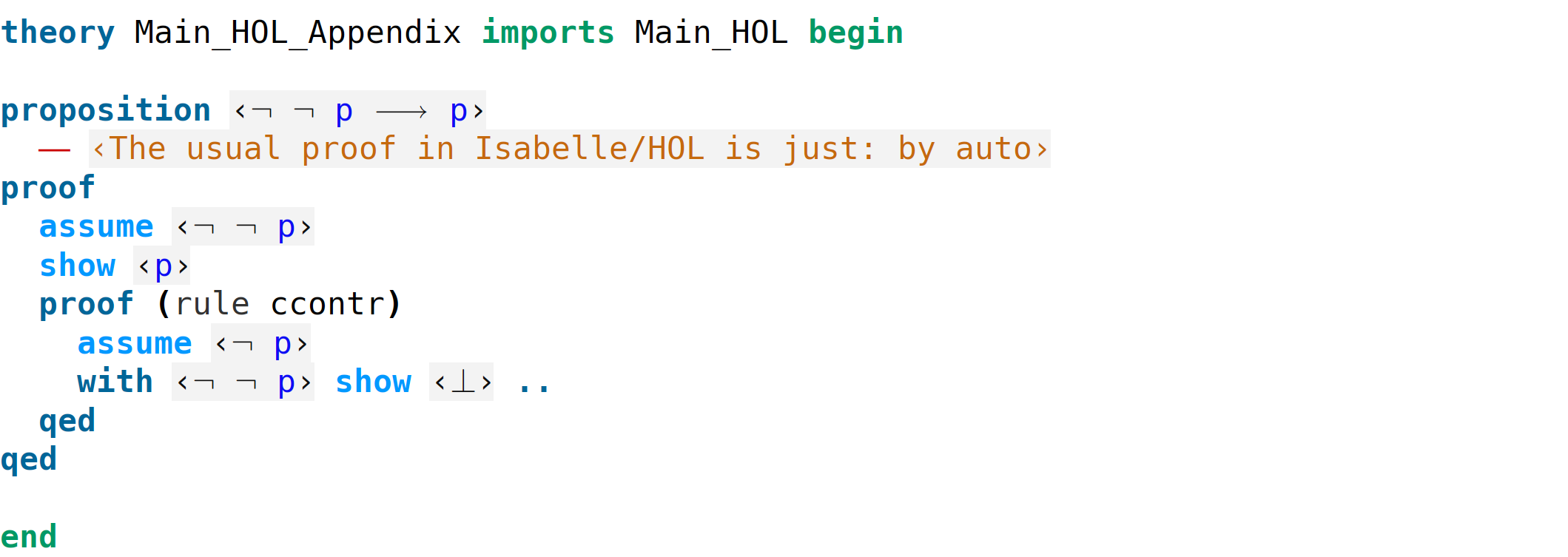}
\caption{A proof from Fig.~\ref{fig:Classical-NaDeA-Appendix} using the pre-defined \isa{ccontr} in Isabelle/HOL.}%
\label{fig:Main-HOL-Appendix}
\end{figure}

In Fig.~\ref{fig:Main-HOL-Appendix} we consider classical contradiction (\isa{ccontr}) already available in Isabelle/HOL by repeating the proof from Fig.~\ref{fig:Classical-NaDeA-Appendix} in this setting.

As exercises we offer a number of formulas to be proved in intuitionistic or classical logic.

\section{The Role of Proof Assistants}
\label{ped-ben}

We have already mentioned a number of features of our course which we feel are pedagogically useful.
In this section we want to stand back, take a more general perspective on our goals, and speculate a little about the future role of Isabelle/Pure (and other proof assistants) in the teaching of logic and mathematics.

Intuitionistic logic, perhaps more than any other logic, is interesting both philosophically and computationally.
The key ideas of intuitionism are due to the Dutch mathematician Luitzen Egbertus Jan Brouwer, and they played a prominent role in the early 20th century debate on the foundations of mathematics.
Brouwer was opposed to Russell's logicism (Brouwer viewed logic as part of mathematics, not as a foundation for it) and also to Hilbert's formalism (intuitionistic logic was formalized not by Brouwer himself but by later followers such as Heyting).
Key themes driving Brouwer's philosophy of mathematics were the activity of proof, and the construction of mathematical objects; these were conceived as mental activities of the ``creative subject'', thus Brouwer attempted to side-step Platonism as well (the view that mathematics is the discovery of truths about some abstract, pre-existing metaphysical realm).
As is well known, variants of Brouwer’s ideas have led to the developments of new logic(s) and have become a source of inspiration for later developments in constructive mathematics and program development.
For more background on the technical, conceptual and historical underpinnings of intuitionism, see \cite{sep-logic-intuitionistic,sep-intuitionism,sep-brouwer}.

To state simply what we shall spell out in the following paragraphs: we believe that teaching intuitionistic logic early in a student's career --- and carefully drawing attention to the differences with classical logic --- is pedagogically useful because it sensitizes students to the nuances of proof. Moreover, we believe that
doing so in the setting of Isabelle/Pure (or indeed, some other proof assistant) forces students to grapple more deeply with these distinctions, enables active experimentation, and prepares them for a future in which technology is likely to play an increasingly important role in logic and mathematics. Let us now try and explain why.

The constructive nature of intuitionistic proof is partially clarified in the BHK-interpretation (the Brouwer, Heyting, Kolmogorov interpretation) which contains clauses like the following:
\begin{quote}
A proof of $\varphi\rightarrow \psi$ is any \textit{construction} which \textit{converts} any proof of $\varphi$ into a proof of $\psi$.
\end{quote}

Similarly: 
\begin{quote}
A negated formula $\neg\varphi$ is proved by giving a \textit{construction} that \textit{converts}
a proof of $\varphi$ into a proof of $\bot$ (where $\bot$ is unprovable).
\end{quote}

The BHK-interpretation is an intuitive interpretation rather than a formal one, because it does not make clear what a ``construction’’ (or ``converts'') mean.
But it is certainly suggestive. For example, any currently unproved mathematical statement, such as the Goldbach Conjecture (every even number greater than two is the sum of two primes) for which we have neither a proof, nor a proof of its negation, is a counterexample to LEM.
Of course, we may lose some LEM counterexamples over time: Fermat’s Last Theorem used to be such a counterexample, but it stopped being one around September 1994, when Andrew Wiles had the crucial insight that lead him to the correct proof.
That is, Brouwer’s view of mathematics emphasizes the dynamic role of creative construction, and the BHK-interpretation captures some of this. Furthermore, the words commonly used to describe the BHK-interpretation, like ``construction’’, ``converts’’ or ``transform’’ have a computational flavour. Indeed, one way of thinking about constructions is as programs: a proof of $\varphi\rightarrow\psi$ is a program that takes a proof of $\varphi$ as input and outputs a proof of $\psi$. 
This way of thinking is commonplace nowadays in various subfields of computer science, such as functional programming, and lies at the heart of what has become known as the Curry-Howard isomorphism.

The late Michael Dummett, an influential English philosopher, published a well-known paper in 1975 (see \cite{dummett1975philosophical}, and also \cite{dummett2000elements}) that attached philosophical significance to intuitionistic logic, but for reasons rather different from those emphasized by Brouwer and his followers.
Brouwer emphasized the role of the ``creative subject’’ – mathematics was some sort of non-linguistic mental activity.
Dummett, on the other hand, strongly emphasizes the linguistic elements of mathematics; see \cite{sep-intuitionism} for a clear discussion.
Dummett emphasizes that mathematics is a language that is to be \textit{used} to be understand.
Mathematics is about learning to \textit{do} certain things -- finding square roots, solving equations, differentiating functions, or proving theorems and corollaries.
Learning the underlying logic of mathematics is all about learning what we need to do with logical statements --- it's about \textit{process} --- and the ability to recognize proofs and to appreciate their fine structure is a key part of this.

And this is the point where we feel the use of proof assistants such as Isabelle/Pure come in: \textit{they open up new research and pedagogical spaces for this process}.
Dummett’s paper is now almost half a century old, thus it precedes the era of interactive technologies which boost human cognitive capacities.
Proof assistants are precisely such a technology: they open up new ways to explore the mathematical landscape via human-machine interaction.
That is: what strikes us as interesting about Dummett's work is the emphasis it places on linguistic/logical explorations underlying mathematics, and the activity of learning ways of proving. And proof assistants provide a truly novel way of zooming in on these activities.

Dummett may or may not be right to single out intuitionistic logic as special --- we prefer to think in terms of \textit{logics} of mathematics.
But as a pedagogical tool, intuitionistic logic is undeniably useful.
Teaching intuitionistic propositional logic early primes students to be aware of the nuances of mathematical proof.
Doing so with the aid of a proof assistant, where they can actively compare classical and intuitionistic logics, amplifies the insights.
It seems plausible that proof assistants will play an increasingly significant role in the development of mathematics and logic, turning research in new and unexpected directions.
We should be training students in ways that will help them explore these new worlds, and our courses can be viewed as simple experiments in how to do this. 

\section{Concluding Remarks}
\label{conc}

We have shown how to use Isabelle/Pure as a logical framework for teaching intuitionistic and classical propositional logic.
Instead of teaching each logic separately, we have shown first how to axiomatize an intuitionistic base and then how to add classical reasoning on top of it.
Along these lines, we want students to learn a framework for describing natural deduction rules instead of only learning the rules themselves.
We see this approach scaling from propositional logic to first-order logic and even further to higher-order logic.
By defining each logic in the same logical framework we hope to make the distinctions between them clear to the students and want to explore this in future work.

\section*{Acknowledgements}

Thanks to Frederik Krogsdal Jacobsen and the anonymous reviewers for comments.

\bibliographystyle{eptcs}
\bibliography{references}

\begin{thebibliography}{10}
\providecommand{\bibitemdeclare}[2]{}
\providecommand{\surnamestart}{}
\providecommand{\surnameend}{}
\providecommand{\urlprefix}{Available at }
\providecommand{\url}[1]{\texttt{#1}}
\providecommand{\href}[2]{\texttt{#2}}
\providecommand{\urlalt}[2]{\href{#1}{#2}}
\providecommand{\doi}[1]{doi:\urlalt{http://dx.doi.org/#1}{#1}}
\providecommand{\eprint}[1]{arXiv:\urlalt{https://arxiv.org/abs/#1}{#1}}
\providecommand{\bibinfo}[2]{#2}

\bibitemdeclare{incollection}{sep-brouwer}
\bibitem{sep-brouwer}
\bibinfo{author}{Mark \surnamestart van Atten\surnameend}
  (\bibinfo{year}{2020}): \emph{\bibinfo{title}{{Luitzen Egbertus Jan
  Brouwer}}}.
\newblock In \bibinfo{editor}{Edward~N. \surnamestart Zalta\surnameend},
  editor: {\sl \bibinfo{booktitle}{The {Stanford} Encyclopedia of Philosophy}},
  \bibinfo{edition}{{S}pring 2020} edition, \bibinfo{publisher}{Metaphysics
  Research Lab, Stanford University}.

\bibitemdeclare{article}{BasinMV98}
\bibitem{BasinMV98}
\bibinfo{author}{David~A. \surnamestart Basin\surnameend},
  \bibinfo{author}{Se{\'{a}}n \surnamestart Matthews\surnameend} \&
  \bibinfo{author}{Luca \surnamestart Vigan{\`{o}}\surnameend}
  (\bibinfo{year}{1998}): \emph{\bibinfo{title}{Labelled Modal Logics:
  Quantifiers}}.
\newblock {\sl \bibinfo{journal}{Journal of Logic, Language and Information}}
  \bibinfo{volume}{7}(\bibinfo{number}{3}), pp. \bibinfo{pages}{237--263},
  \doi{10.1023/A:1008278803780}.

\bibitemdeclare{inproceedings}{Isar08}
\bibitem{Isar08}
\bibinfo{author}{Stefan \surnamestart Berghofer\surnameend} \&
  \bibinfo{author}{Makarius \surnamestart Wenzel\surnameend}
  (\bibinfo{year}{2008}): \emph{\bibinfo{title}{Logic-Free Reasoning in
  {Isabelle/Isar}}}.
\newblock In \bibinfo{editor}{Serge \surnamestart Autexier\surnameend},
  \bibinfo{editor}{John~A. \surnamestart Campbell\surnameend},
  \bibinfo{editor}{Julio \surnamestart Rubio\surnameend},
  \bibinfo{editor}{Volker \surnamestart Sorge\surnameend},
  \bibinfo{editor}{Masakazu \surnamestart Suzuki\surnameend} \&
  \bibinfo{editor}{Freek \surnamestart Wiedijk\surnameend}, editors: {\sl
  \bibinfo{booktitle}{Intelligent Computer Mathematics, 9th International
  Conference, {AISC} 2008, 15th Symposium, Calculemus 2008, 7th International
  Conference, {MKM} 2008, Birmingham, UK, July 28 - August 1, 2008.
  Proceedings}}, {\sl \bibinfo{series}{Lecture Notes in Computer Science}}
  \bibinfo{volume}{5144}, \bibinfo{publisher}{Springer}, pp.
  \bibinfo{pages}{355--369}, \doi{10.1007/978-3-540-85110-3\_31}.

\bibitemdeclare{article}{DionisioGM05}
\bibitem{DionisioGM05}
\bibinfo{author}{F.~Miguel \surnamestart Dion\'{i}sio\surnameend},
  \bibinfo{author}{Paula \surnamestart Gouveia\surnameend} \&
  \bibinfo{author}{Jo{\~{a}}o \surnamestart Marcos\surnameend}
  (\bibinfo{year}{2005}): \emph{\bibinfo{title}{Defining and using deductive
  systems with {Isabelle}}}.
\newblock {\sl \bibinfo{journal}{Computing, Philosophy, and Cognition}}, pp.
  \bibinfo{pages}{271--293}.

\bibitemdeclare{incollection}{dummett1975philosophical}
\bibitem{dummett1975philosophical}
\bibinfo{author}{Michael \surnamestart Dummett\surnameend}
  (\bibinfo{year}{1975}): \emph{\bibinfo{title}{The philosophical basis of
  intuitionistic logic}}.
\newblock In: {\sl \bibinfo{booktitle}{Studies in Logic and the Foundations of
  Mathematics}}, \bibinfo{volume}{80}, \bibinfo{publisher}{Elsevier}, pp.
  \bibinfo{pages}{5--40}.

\bibitemdeclare{book}{dummett2000elements}
\bibitem{dummett2000elements}
\bibinfo{author}{Michael \surnamestart Dummett\surnameend}
  (\bibinfo{year}{2000}): \emph{\bibinfo{title}{Elements of intuitionism}}.
\newblock \bibinfo{publisher}{Oxford University Press}.

\bibitemdeclare{inproceedings}{ThEdu19}
\bibitem{ThEdu19}
\bibinfo{author}{Asta~Halkj{\ae}r \surnamestart From\surnameend},
  \bibinfo{author}{Alexander~Birch \surnamestart Jensen\surnameend},
  \bibinfo{author}{Anders \surnamestart Schlichtkrull\surnameend} \&
  \bibinfo{author}{J{\o}rgen \surnamestart Villadsen\surnameend}
  (\bibinfo{year}{2019}): \emph{\bibinfo{title}{Teaching a Formalized Logical
  Calculus}}.
\newblock In \bibinfo{editor}{Pedro \surnamestart Quaresma\surnameend},
  \bibinfo{editor}{Walther \surnamestart Neuper\surnameend} \&
  \bibinfo{editor}{Jo{\~{a}}o \surnamestart Marcos\surnameend}, editors: {\sl
  \bibinfo{booktitle}{Proceedings 8th International Workshop on Theorem Proving
  Components for Educational Software, ThEdu@CADE 2019, Natal, Brazil, 25th
  August 2019}}, {\sl \bibinfo{series}{{EPTCS}}} \bibinfo{volume}{313}, pp.
  \bibinfo{pages}{73--92}, \doi{10.4204/EPTCS.313.5}.

\bibitemdeclare{inproceedings}{ThEdu20}
\bibitem{ThEdu20}
\bibinfo{author}{Asta~Halkj{\ae}r \surnamestart From\surnameend},
  \bibinfo{author}{J{\o}rgen \surnamestart Villadsen\surnameend} \&
  \bibinfo{author}{Patrick \surnamestart Blackburn\surnameend}
  (\bibinfo{year}{2020}): \emph{\bibinfo{title}{Isabelle/{HOL} as a
  Meta-Language for Teaching Logic}}.
\newblock In \bibinfo{editor}{Pedro \surnamestart Quaresma\surnameend},
  \bibinfo{editor}{Walther \surnamestart Neuper\surnameend} \&
  \bibinfo{editor}{Jo{\~{a}}o \surnamestart Marcos\surnameend}, editors: {\sl
  \bibinfo{booktitle}{Proceedings 9th International Workshop on Theorem Proving
  Components for Educational Software, ThEdu@IJCAR 2020, Paris, France, 29th
  June 2020}}, {\sl \bibinfo{series}{{EPTCS}}} \bibinfo{volume}{328}, pp.
  \bibinfo{pages}{18--34}, \doi{10.4204/EPTCS.328.2}.

\bibitemdeclare{article}{FOL-Axiomatic-AFP}
\bibitem{FOL-Axiomatic-AFP}
\bibinfo{author}{Asta~Halkjær \surnamestart From\surnameend}
  (\bibinfo{year}{2021}): \emph{\bibinfo{title}{Soundness and Completeness of
  an Axiomatic System for First-Order Logic}}.
\newblock {\sl \bibinfo{journal}{Archive of Formal Proofs}}.
\newblock \bibinfo{note}{\url{https://isa-afp.org/entries/FOL_Axiomatic.html},
  Formal proof development}.

\bibitemdeclare{incollection}{sep-intuitionism}
\bibitem{sep-intuitionism}
\bibinfo{author}{Rosalie \surnamestart Iemhoff\surnameend}
  (\bibinfo{year}{2020}): \emph{\bibinfo{title}{{Intuitionism in the Philosophy
  of Mathematics}}}.
\newblock In \bibinfo{editor}{Edward~N. \surnamestart Zalta\surnameend},
  editor: {\sl \bibinfo{booktitle}{The {Stanford} Encyclopedia of Philosophy}},
  \bibinfo{edition}{{F}all 2020} edition, \bibinfo{publisher}{Metaphysics
  Research Lab, Stanford University}.

\bibitemdeclare{incollection}{sep-logic-intuitionistic}
\bibitem{sep-logic-intuitionistic}
\bibinfo{author}{Joan \surnamestart Moschovakis\surnameend}
  (\bibinfo{year}{2021}): \emph{\bibinfo{title}{{Intuitionistic Logic}}}.
\newblock In \bibinfo{editor}{Edward~N. \surnamestart Zalta\surnameend},
  editor: {\sl \bibinfo{booktitle}{The {Stanford} Encyclopedia of Philosophy}},
  \bibinfo{edition}{{F}all 2021} edition, \bibinfo{publisher}{Metaphysics
  Research Lab, Stanford University}.

\bibitemdeclare{book}{nipkow+02}
\bibitem{nipkow+02}
\bibinfo{author}{Tobias \surnamestart Nipkow\surnameend},
  \bibinfo{author}{Lawrence~C. \surnamestart Paulson\surnameend} \&
  \bibinfo{author}{Markus \surnamestart Wenzel\surnameend}
  (\bibinfo{year}{2002}): \emph{\bibinfo{title}{{Isabelle/HOL} --- A Proof
  Assistant for Higher-Order Logic}}.
\newblock {\sl \bibinfo{series}{Lecture Notes in Computer Science}}
  \bibinfo{volume}{2283}, \bibinfo{publisher}{Springer},
  \doi{10.1007/3-540-45949-9}.

\bibitemdeclare{article}{Paulson89}
\bibitem{Paulson89}
\bibinfo{author}{Lawrence~C. \surnamestart Paulson\surnameend}
  (\bibinfo{year}{1989}): \emph{\bibinfo{title}{The Foundation of a Generic
  Theorem Prover}}.
\newblock {\sl \bibinfo{journal}{J. Autom. Reason.}}
  \bibinfo{volume}{5}(\bibinfo{number}{3}), pp. \bibinfo{pages}{363--397},
  \doi{10.1007/BF00248324}.

\bibitemdeclare{inproceedings}{ThEdu18}
\bibitem{ThEdu18}
\bibinfo{author}{Anders \surnamestart Schlichtkrull\surnameend},
  \bibinfo{author}{J{\o}rgen \surnamestart Villadsen\surnameend} \&
  \bibinfo{author}{Andreas~Halkj{\ae}r \surnamestart From\surnameend}
  (\bibinfo{year}{2018}): \emph{\bibinfo{title}{Students' Proof Assistant
  {(SPA)}}}.
\newblock In \bibinfo{editor}{Pedro \surnamestart Quaresma\surnameend} \&
  \bibinfo{editor}{Walther \surnamestart Neuper\surnameend}, editors: {\sl
  \bibinfo{booktitle}{Proceedings 7th International Workshop on Theorem proving
  components for Educational software, THedu@FLoC 2018, Oxford, United Kingdom,
  18 july 2018}}, {\sl \bibinfo{series}{{EPTCS}}} \bibinfo{volume}{290}, pp.
  \bibinfo{pages}{1--13}, \doi{10.4204/EPTCS.290.1}.

\bibitemdeclare{inproceedings}{ThEdu17}
\bibitem{ThEdu17}
\bibinfo{author}{J{\o}rgen \surnamestart Villadsen\surnameend},
  \bibinfo{author}{Andreas~Halkj{\ae}r \surnamestart From\surnameend} \&
  \bibinfo{author}{Anders \surnamestart Schlichtkrull\surnameend}
  (\bibinfo{year}{2017}): \emph{\bibinfo{title}{Natural Deduction and the
  {Isabelle} Proof Assistant}}.
\newblock In \bibinfo{editor}{Pedro \surnamestart Quaresma\surnameend} \&
  \bibinfo{editor}{Walther \surnamestart Neuper\surnameend}, editors: {\sl
  \bibinfo{booktitle}{Proceedings 6th International Workshop on Theorem proving
  components for Educational software, ThEdu@CADE 2017, Gothenburg, Sweden, 6
  Aug 2017}}, {\sl \bibinfo{series}{{EPTCS}}} \bibinfo{volume}{267}, pp.
  \bibinfo{pages}{140--155}, \doi{10.4204/EPTCS.267.9}.

\bibitemdeclare{inproceedings}{PAAR}
\bibitem{PAAR}
\bibinfo{author}{J{\o}rgen \surnamestart Villadsen\surnameend},
  \bibinfo{author}{Anders \surnamestart Schlichtkrull\surnameend} \&
  \bibinfo{author}{Andreas~Halkj{\ae}r \surnamestart From\surnameend}
  (\bibinfo{year}{2018}): \emph{\bibinfo{title}{A Verified Simple Prover for
  First-Order Logic}}.
\newblock In \bibinfo{editor}{Boris \surnamestart Konev\surnameend},
  \bibinfo{editor}{Josef \surnamestart Urban\surnameend} \&
  \bibinfo{editor}{Philipp \surnamestart R{\"{u}}mmer\surnameend}, editors:
  {\sl \bibinfo{booktitle}{Proceedings of the 6th Workshop on Practical Aspects
  of Automated Reasoning (PAAR 2018) co-located with Federated Logic Conference
  2018 (FLoC 2018), Oxford, UK, 19 July 2018}}, {\sl \bibinfo{series}{{CEUR}
  Workshop Proceedings}} \bibinfo{volume}{2162},
  \bibinfo{publisher}{CEUR-WS.org}, pp. \bibinfo{pages}{88--104}.
\newblock \urlprefix\url{http://ceur-ws.org/Vol-2162/paper-08.pdf}.

\bibitemdeclare{inproceedings}{Pure08}
\bibitem{Pure08}
\bibinfo{author}{Makarius \surnamestart Wenzel\surnameend},
  \bibinfo{author}{Lawrence~C. \surnamestart Paulson\surnameend} \&
  \bibinfo{author}{Tobias \surnamestart Nipkow\surnameend}
  (\bibinfo{year}{2008}): \emph{\bibinfo{title}{The {Isabelle} Framework}}.
\newblock In \bibinfo{editor}{Otmane~A{\"{\i}}t \surnamestart
  Mohamed\surnameend}, \bibinfo{editor}{C{\'{e}}sar~A. \surnamestart
  Mu{\~{n}}oz\surnameend} \& \bibinfo{editor}{Sofi{\`{e}}ne \surnamestart
  Tahar\surnameend}, editors: {\sl \bibinfo{booktitle}{Theorem Proving in
  Higher Order Logics, 21st International Conference, TPHOLs 2008, Montreal,
  Canada, August 18-21, 2008. Proceedings}}, {\sl \bibinfo{series}{Lecture
  Notes in Computer Science}} \bibinfo{volume}{5170},
  \bibinfo{publisher}{Springer}, pp. \bibinfo{pages}{33--38},
  \doi{10.1007/978-3-540-71067-7\_7}.

\end{thebibliography}

\end{document}